\documentclass[acmjacm]{article}
\usepackage{amssymb,amsmath,shortvrb,psfrag,graphicx,subfigure,alltt,fullpage,longtable,comment,capt-of,rotating}

\providecommand{\OO}[1]{\mathop{\mathrm{O}}\bigl(#1\bigr)}
\def\reffig#1{Fig.\,\ref{#1}}
\def\reftable#1{Table \,\ref{#1}}
\def\ltcenter#1{#1}
\def\refsec#1{Section \,\ref{#1}}

\begin{document}
\title{Towards Parallel Computing on the Internet: Applications, Architectures, Models and Programming Tools}
\author{Elankovan Sundararajan and Aaron Harwood \\
Department of Computer Science and Software Engineering, \\
The University of Melbourne, \\
Carlton 3053, Victoria Australia.\\
Email:\{esund,aharwood\}@csse.unimelb.edu.au.}

\date{}

\maketitle

\tableofcontents \setcounter{page}{1}
\begin{abstract}
The development of Internet wide resources for general purpose
parallel computing poses the challenging task of matching
computation and communication complexity. A number of parallel
computing models exist that address this for traditional parallel
architectures, and there are a number of emerging models that
attempt to do this for large scale Internet-based systems like
computational grids. In this survey we cover the three fundamental
aspects -- application, architecture and model, and we show how
they have been developed over the last decade. We also cover
programming tools that are currently being used for parallel
programming in computational grids. The trend in conventional
computational models are to put emphasis on efficient
communication between participating nodes by adapting different
types of communication to network conditions. Effects of dynamism
and uncertainties that arise in large scale systems are evidently
important to understand and yet there is currently little work
that addresses this from a parallel computing perspective.
\end{abstract}

\section{Introduction}\label{sec1}
The field of High Performance Computing (HPC) has evolved to include a
variety of very complex architectures, computing models and problem solving
environments. HPC architectures consist of Massively Parallel Processors
(MPPs), clusters and constellation architectures and they typically use
hundreds to hundreds of thousands of CPUs.  Some application problems involve
large real time data that must be processed as soon as possible, while
others involve a high degree of computational complexity. Computing models on
the other hand, provide a bridge between hardware and software to assist
application developers in designing and writing parallel applications that
efficiently utilize the available parallel architecture. Problem solving
environments provide comprehensive computational facilities for programmers
to develop parallel applications on these platforms. These environments
usually consists of programming tools, utilities, libraries, debuggers,
profilers, etc.

The extent to which a system can be called a HPC architecture is
relatively ambiguous and dynamic, because the contemporary HPC
architecture and notion of HPC can be liberally extended to cover
collections of resources that are combined to solve a single
problem. These definitions lead us to consider computational
grids~\cite{60} as (commodity) supercomputers and indeed
computational grids are being used to solve problems that were and
still are sometimes solved by the classical HPC architectures. In
general, it is clear that problems are migrating from classical
HPC architectures towards the contemporary computational grid (or
at least that the use of the Internet is becoming prevalent in
order to tie more computing resources together), either explicitly
by direct programming efforts or implicitly through
virtualization. Some problems are harder than others to migrate
and this survey covers the approaches that have and are being used
to overcome the associated difficulties.

Developing applications for HPC is not comparable to developing applications
for a single processor mainly because of the complexity involved in the HPC
architectures. The challenge that this survey addresses is how the application
developer can understand the differences in complexity between the problem and
communication imposed by the architecture. By surveying the past and present
computational models and in particular those that are associated with
computational grids we provide a resource for future parallel programmers to
better understand the ways in which the computational grid architecture affects
their programs.  A model allows the determination of computational and
communication complexities associated with a given problem, as expressed by the
hardware. It plays an important role to reflect the salient computing
characteristics of a particular architecture to develop fast and efficient
algorithms and provides information on the performance of an application.

When developing application software for HPC, parallel application
developers must emphasize both extreme ends of the architecture,
namely the memory hierarchy and the inter-processor communication.
This is due to the cost associated in accessing large data sets.
Furthermore, the rate of data access is not as fast as the rate of
computation performed by processors due to bandwidth limitation
for both the inter-processor and processor-memory data transfer.
All of the emerging models therefore consider the data movement
costs in a system under consideration, as accurately as possible.
It is also important to note that a model may provide good
representation of an architecture, but to gauge an application's
performance it is necessary to take into consideration how
efficiently the application can be implemented (efficiency of
coding).

\begin{figure}[htbp]
\begin{minipage}[b]{0.48\columnwidth}%
    \centering
    \includegraphics[height=2.0in,width=2.5in,angle=0]{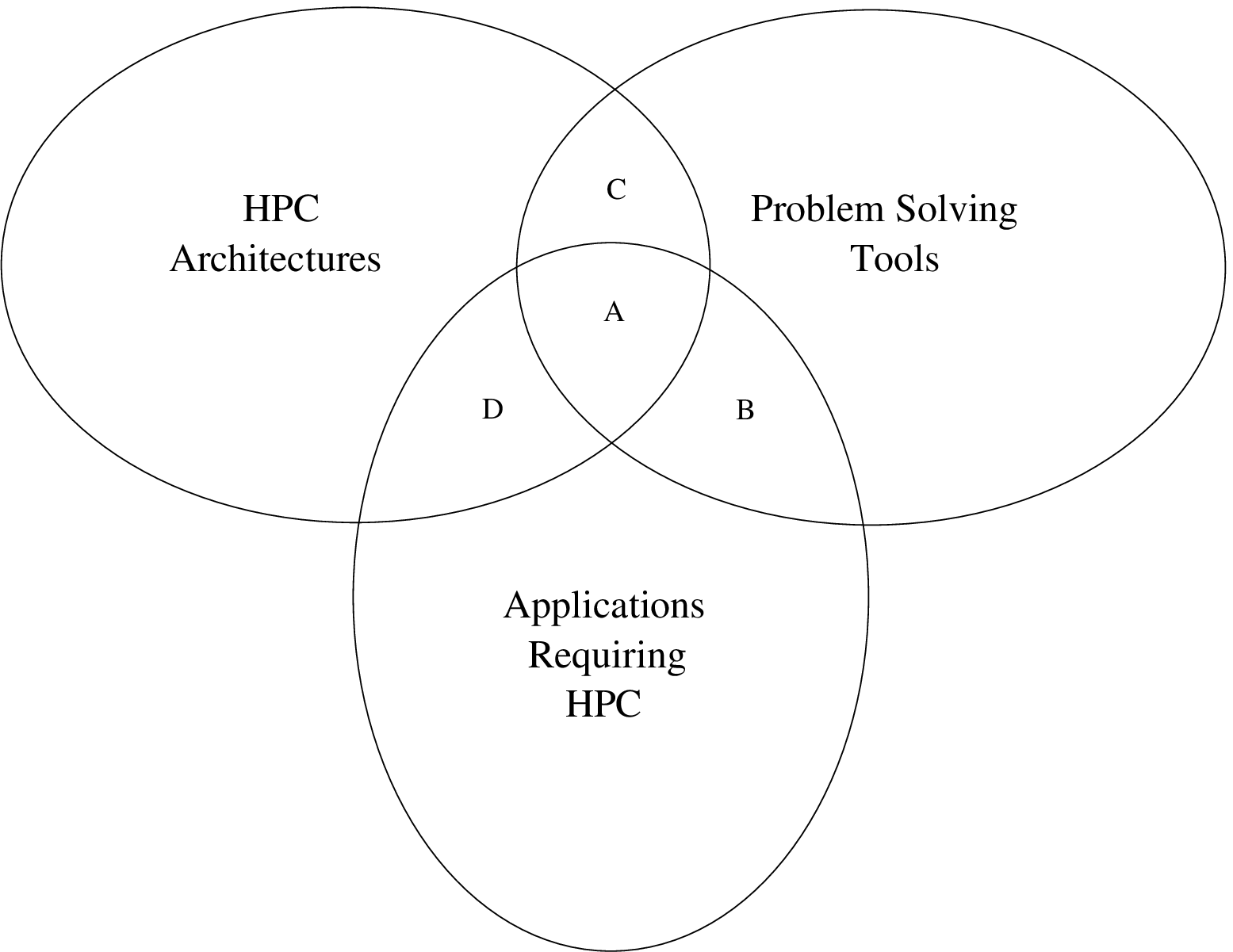}
    \caption{\label{Prob_arch_model_Plang_code} The
    relationship between HPC architectures, problem solving tools and applications requiring HPC.}
\end{minipage}%
\hfill%
\begin{minipage}[b]{0.48\columnwidth}%
\begin{tabular}{p{0.7in}|p{2in}}
\hline
Overlapping Region. & Description.\\
\hline
A   &   Computational model providing information on performance of parallel programs.\\
\hline
B   &   Algorithm parameters (e.g data size, communication type, computational complexity, etc.) and problem solving tools.\\
\hline
C   &   Performance tuning parameters (e.g. number of processors, latency, bandwidth, shared/distributed memory, etc.).\\
\hline
D   &   Requirements for solving problem in reasonable amount of time (e.g. storage, memory \& computational capacity, number of processors and algorithms).\\
\hline
\end{tabular}
\captionof{table}{Explanation for the overlapping region in
\reffig{Prob_arch_model_Plang_code}.}
\end{minipage}
\end{figure}

Relationships between HPC architectures, problem solving tools, and
applications requiring HPC are shown in \reffig{Prob_arch_model_Plang_code}.
The overlapping region A, depicts the computational performance of a parallel
program, region B shows the use of problem solving tools and algorithms to
solve the problem without considering the parallel architecture, region C
represents performance tuning parameters with information from parallel
architecture, and region D represents algorithms and the requirements for
solving the problem in a reasonable amount of time. HPC architectures and grand
challenge problems decide which type of model should be used and in turn the
model decides parameters to be used in the programming language.

\subsection{Objective}
The main objective of this paper is to show the importance of an
accurate computational model in solving large scale application on
HPC architectures. We begin by looking at some of the applications
that require HPC, the characteristics of these applications such
as memory requirements, computational requirements, storage space,
communication and computational complexity, and algorithms
required to solve this problem. Later, we look at the
characteristics of architectures that have evolved to attempt to
solve these application as fast as possible. Here we list some of
the important characteristics of these architectures. The
motivation for new HPC architectures are the challenges introduced
by the large scale problems, while the motivation for
computational models are to efficiently solve the problems on the
available architecture. Some architectures are more suitable for
certain types and sizes of problems, and it is important to have
an idea beforehand on the suitability of the architecture before
the problem is solved on it. This is where the computational model
will play its role as a bridge between them. Hence, we study some
of the more popular parallel computational models that have been
used in the past and also look at some of the conventional
computational models. It becomes clear that the new models are
moving towards the direction of assisting adaptation of parallel
computing softwares to the dynamic behavior of the architecture.

\subsection{Organization}
We divide this paper into six main sections. In \refsec{sec2}, we
look at different applications that require the use of HPC
architectures. We list some significant characteristics of these
applications that highlights the configuration requirement for
HPC. Next, in \refsec{sec3}, we briefly look at recent HPC
architectures. Here we list some of the important properties of
these architectures. This is important to measure how the parallel
computing model has evolved to better reflect HPC architectures.
\refsec{Computational_models}, looks at traditional parallel
computing models and conventional parallel models used to design
parallel algorithm and predict performance of HPC architectures.
In this section, we investigate factors considered by different
parallel models that have been developed and look at how the
development in architectures have influenced the models. We also
discuss some parallel computing models that are developed for Grid
environment. \refsec{sec6}, discusses some of the popular parallel
programming libraries used by HPC communities for both traditional
supercomputers and also the Grid. \refsec{sec7}, concludes the
paper and provides suggestion on attributes that should be
considered for parallel computing model on Grid environment.

\section{Applications challenges}\label{sec2}
In this section, we describe the ever increasing need for HPC
facilities and we give insight into the computational complexities
and other demands of a number of applications in the field of
computational science; which is useful for identifying the required
HPC facilities and computational models.

Many fields in science and engineering have computationally
intensive problems that are intractable without the use of
HPC. Most of these problems come under the category of
computational sciences. Problems such as climate modeling (which
consists of atmosphere model, ocean model, hurricane model,
hydrological model and sea-ice model), plasma physics (to produce
safe, clean and cost-effective energy from nuclear fusion),
engineering design (of aircraft, ships, and vehicles),
bio-informatics and computational biology, geophysical exploration
and geoscience, astrophysics, material science and nanotechnology,
defense (cracking cryptography code), computational fluid
dynamics, and computational physics are computationally demanding.
The characteristics of these applications listed in
\reftable{app_char} are:

\begin{description}
    \item[Memory requirement] The size of main memory required to store data
for computation. This measurement is important for selection of
suitable computing resources. Resources with memory less than this
threshold will deteriorate the application performance as more
time will be required to access data from secondary storage.
    \item[Computational requirement] The amount of Floating Point Operations per Second (FLOPS) required to undertake the complexity of the
problem in a ``reasonable amount of time" as some application
involves real-time data. This measure depends on several factors
such as abstraction of the problem and the size of computation.
    \item[Storage] The minimum amount of storage space required by the application
to store simulation results for visualization purposes or to
store sufficient amount of data to be used in
computation for ``reasonable amount of accuracy". This value will be useful to chose resources that meet
the requirement and avoid loss of information.
    \item[Communication complexity] Is the amount of information that needs to be
    communicated between computing nodes to successfully complete a computation.
     This provides information on the communication needs of an algorithm for executing across multiple computing nodes.
     It is in particular important for the purpose of selecting optimal number of resources to use for a particular problem size.
    \item[Computational complexity] This gives information on how the
complexity of an algorithm grows as the size of the problem
increases. This information is critical for choosing appropriate
computing resources.
    \item[Algorithms] Different types of algorithms that can be used to
solve a particular problem.
\end{description}

A typical problem of computational science involves finding the
solution to models of real world phenomenon. Many of these models
use Partial Differential Equations (PDEs) and are approximated
using discretized equations. For better approximation, higher
resolution must be used and this demands more computational power.
All of these grand challenge problems are difficult to be solve
efficiently with better accuracy due to a number of reasons:
\emph{1)} Limitation in capability of hardwares, \emph{2)}
Algorithms used to solve the problems and \emph{3)} Tools that are
available for a programmer to solve these problems and analyze the
results. The term ``Grand Challenge" used in previous statement
was coined by Nobel Laureate Kenneth G. Wilson, who also
articulated the current concept of ``computational science" as a
third way of doing science~\cite{Gustafson94}. The Grand Challenge
problems have the following properties in common: \emph{1)} They
are questions to which many scientists and engineers would like to
know answers; \emph{2)} They are difficult and it is not known how
to do them right now; \emph{3)} It may be done using computers but
the current computers are not fast enough.~\cite{Gustafson94}

\begin{figure}[htbp]
\psfrag{Data Minning}{Data mining}
\centerline{\rotatebox{-90}{\includegraphics[height=5in,width=3.94in]{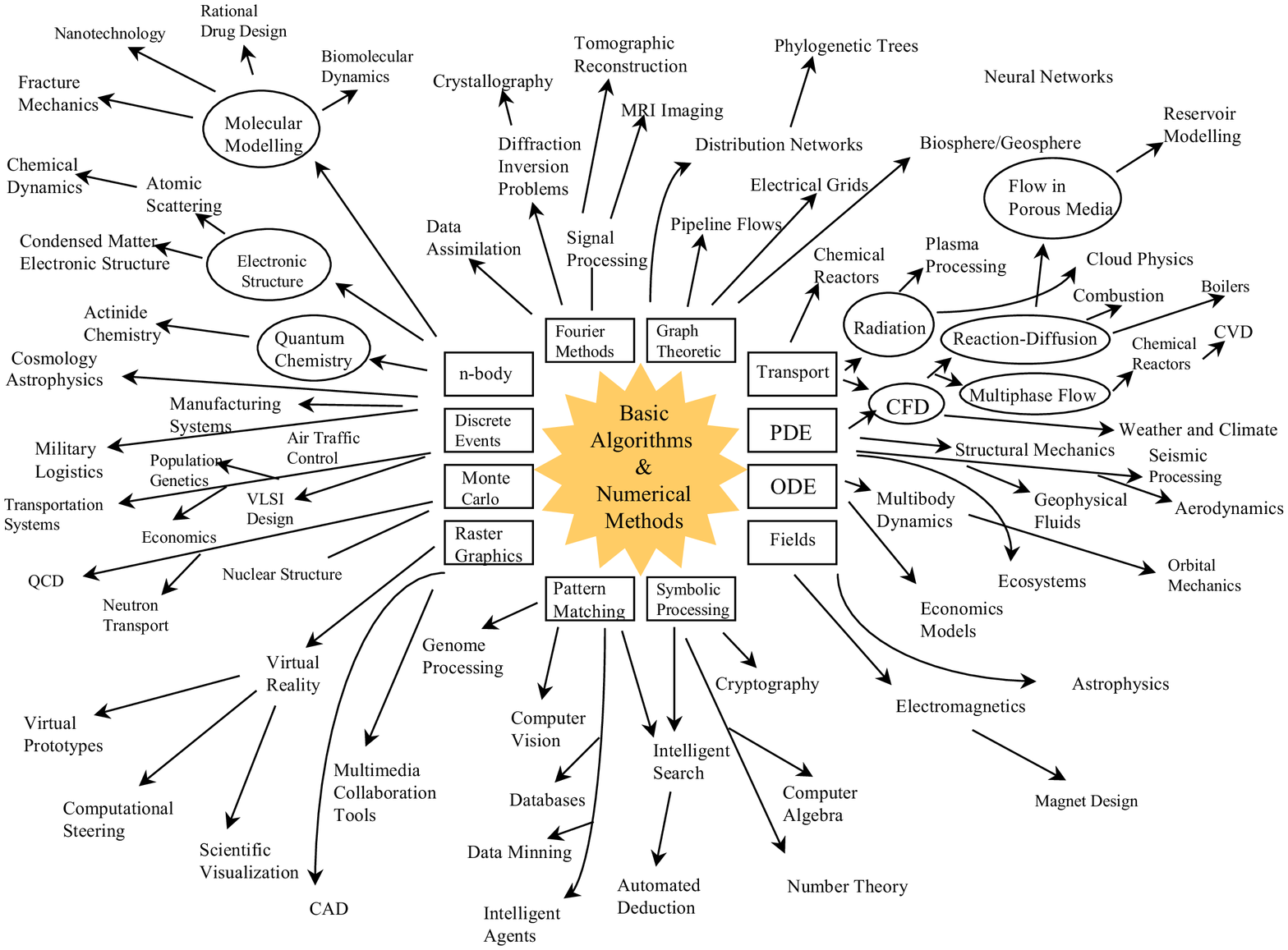}}}
\caption{\label{application} Research areas that require immense
computational power to complement theory and experiment.[Courtesy:
Rick Stevens]}
\end{figure}

Basic algorithms and numerical algorithms play important role in
many computationally intensive scientific applications. Some of
these grand challenge applications and algorithms that are used to
solve them using HPC are depicted in
\reffig{application}~\footnote{http://www.cacr.caltech.edu/pflops2/presentations/stevenspeta2appsintro.pdf}.
It is interesting to observe that all these applications depend on
some of the most fundamental algorithms. Many highly tuned
parallel computational libraries and computational kernels are
available for these algorithms to be used on dedicated computing
platforms. However, they are not proven to be as efficient on
computing resources distributed across the WAN.

\begin{center}
\footnotesize{
\begin{longtable}{p{1.5in}p{0.6in}p{0.7in}p{0.4in}p{1in}p{0.8in}}
\caption{\label{app_char}Characteristics of Grand Challenge
applications.}
 \\
  \hline \\
    \centering Applications & \centering \begin{sideways}Memory\end{sideways} \begin{sideways}requirement\end{sideways}& \centering \begin{sideways}Computational\end{sideways} \begin{sideways}requirement\end{sideways}& \centering \begin{sideways}Storage\end{sideways} &
    \centering \begin{sideways}Communication\end{sideways} \begin{sideways}complexity\end{sideways} &
    \centering \begin{sideways}Computational\end{sideways}
    \begin{sideways}complexity\end{sideways}\tabularnewline
     \hline
  \endhead
   Climate Modeling: Atmosphere model resolution of $75km$ and ocean model resolution of $10km$. & $>1$TB  depending on the resolution of model.& $100$--$150$ TFLOP/s for high resolution and highly complex model & $>23$TB for a single century simulation. & FFT--$\OO{P^2}$ where $P$ is the No. of processors. & $\OO{N^2}$ with $N$--Size of
   resolution.\\\\
   \multicolumn{6}{l}{Algorithms: FFT, Finite Difference, Finite element method.}\\

  \hline

   Bioinformatics and Computational biology. & $>$ Several hundred MB/processor. & $\approx 100$ TFLOP/s--few PFLOP/s. & $>1$PB. & $\OO{P}$--$\OO{P^2}$ where $P$ is the No. of processors. & $\OO{N^2}$--$\OO{N^3}$ where $N$ is the No. of
   atoms.\\\\
   \multicolumn{6}{l}{Algorithms: Complex Combinatorial, Graph Theoretic, Differential Equation Solver.}\\

  \hline

   Astrophysics simulations. &  $>10$TB. &  $\approx100$TFLOP/s--$10$PFLOP/s. &  $>1$PB. & FMM:$\OO{loglog(P)}$ and $\OO{log(P)}$ for balanced and exponential distribution respectively, FFT--$\OO{P^2}$  where $P$ is the No. of processors. &
   $\OO{N^{10}}$--$\OO{N^{15}}$ where $N$ is the size of the problem.\\\\
   \multicolumn{6}{l}{Algorithms: Fast Multipole Method (FMM), Multi-Scale Dense Linear Algebra, Parallel $3$D FFTs, Spherical
   Transforms,}\\
   \multicolumn{6}{l}{Particle methods and adaptive mesh refinement.}\\

  \hline

   Computational material science and Nanoscience. &  & several PFLOP/s   & & FFT--$\OO{P^2}$ where $P$ is the No. of processors.& $\OO{N^3}$--$\OO{N^7}$ with $N$ as No. of atoms in a
   molecule.\\\\
   \multicolumn{6}{l}{Algorithms:
   Quantum Molecular Dynamics (QMD), Quantum Monte Carlo (QMC), Dense Linear
   Algebra,}\\
   \multicolumn{6}{l}{Parallel $3$D FFT, Iterative Eigen Solvers.}\\

  \hline
   Computational Fluid Dynamics (CFD). & $>400$GB for double precision arithmetic. & $1$ PFLOP/s--few PFLOP/s. & $1$TB. &$\OO{P}$--$\OO{P^2}$ where $P$ is the No. of processors
   &$\OO{Nlog(N)}$--$\OO{N^2}$ where $N$ is the size of the problem.\\\\
   \multicolumn{6}{l}{Algorithms: Finite Difference, Finite Element, Finite Volume, Pseudospectral and Spectral methods.}\\

  \hline

   Computational Physics. & & & & & \\
   Plasma science. & $>50$TB & $100$TFLOPs--few PFLOP/s. & $>27$PB & $\OO{P}$--$\OO{P^2}$ where $P$ is the No. of processors.&
   $\OO{N^{10}}$.\\\\
   \multicolumn{6}{l}{Algorithms: Gyrokinetic (GK), Gyro-Landau-fluid (GLF), Nonlinear Solvers, Adaptive Mesh
   Refinement,}\\
   \multicolumn{6}{l}{Dense Linear Algebra and Particle Methods.}\\

  \hline

   Particle Accelerator Simulation. & & & & & \\
   Electron cooling. & $>5$GB. & $\approx10^6$--$10^7$ TFLOPS per run. & $>2$TB & $\OO{P}$--$\OO{P^2}$, where $P$ is the No. of processors. &
   $\OO{NlogN}$--$\OO{N^2}$ \\
   Beam heating. &  $>1$TB. & $\approx 10^3$--$10^4$ TFLOPS per run. & $>2$TB & $\OO{P}$-$\OO{P^2}$, with $P$ is the No. of processors. & $\OO{NlogN}$--$\OO{N^2}$ \\\\
   \multicolumn{6}{l}{Algorithms: Fast Fourier Transform (FFT), Fast Multipole Method (FMM), Finite Difference method (FDM).}\\

   \hline
   Computational chemistry.& & $>1$PFLOP/s.& & FMM: $\OO{loglog(P)}$ and $\OO{log(P)}$ for balanced and exponential distribution respectively, where $P$ is the No. of processors. & CCSD(T): $\OO{N^7}$ where $N$ is the No. of
   electrons.\\\\
   \multicolumn{6}{l}{Algorithms: CCSD(T) method, FMM method.}\\

   \hline

   Combustion science: turbulent reacting flow computation. & $\approx 8$ TB. & $\approx 30$ PFLOP/s. & $\approx25$TB. &$\OO{P}$--$\OO{P^2}$ where $P$ is the No. of processors. & $\OO{N^3}$--$\OO{N^4}$ with $N$ as the reciprocal of the mesh interval and a coefficient reciprocal in Mach
   number.\\\\
   \multicolumn{6}{l}{
   Algorithms: Semi-Implicit Adaptive Meshing, Finite Difference Method, Zero Dimensional
   Physics,FFT,}\\
   \multicolumn{6}{l}{Adaptive Mesh Refinement and Lagrangian Particle Methods.}\\

  \hline

\end{longtable}
}
\end{center}

In this section we discuss some of the grand challenge
applications that require immense computational power for producing higher accuracy in their solution.\\ \\

\subsection{Climate modeling}
Climate models are used to study the dynamics of the weather and
climate system for predicting future climate conditions. The
climate model consists of several important components of climate
systems: an atmosphere model, an ocean model, a hydrological (a
combined land-vegetation-river transport) model, and a sea-ice
model. Some climate models also incorporate chemical cycles such
as carbon, sulfate, methane, and nitrogen cycles. The most
important and least parameterizable influence on climate change is
the response of cloud systems and they are best treated by using
smaller grid sizes of 1km ~\cite{53,54}.
Climate simulations of $100$ to $1000$ years require thousands of
computational hours on supercomputers. However, it is also very
important to note that reaching an equilibrium climate via
simulation requires thousands of years of simulation, further
hundreds of years of simulation to evaluate climate change beyond
equilibrium and tens of runs to determine the envelope of possible
climate changes for a given emission scenario, and a multitude of
scenarios for future emission of greenhouse gases and human
responses to climate change. These extended simulations need the
integration of the nonlinear equations using small time steps of
seconds for probing important phenomena such as internal waves and
convection. Complex climate model with more in-depth physical
behavior can be simulated to refine further the understanding of
the repercussion on climate and to take necessary precautions
~\cite{54}. Climate simulations require a very large memory size
of more than $1$ Terabytes depending on the resolution used and
storage size of more than $23$ Terabytes for a single-century
simulation. Spectral Methods, Finite Difference and Finite Element
Methods are usually used for climate simulations ~\cite{Simon04}.

\subsection{Bioinformatics and Computational biology}
Advancement in computation and information technology has provided
the impetus for future developments in biology and biomedicine.
Understanding how cells and systems of cells function in order to
improve human health, longevity, and to treat diseases in
molecular biology requires immense computing power. The complexity
of molecular systems in terms of number of molecules and type of
molecules contributes to the computational needs. For example,
finding multiple alignments of the sequences of bacterial genomes
can only be attempted with new algorithms using a petaflops
supercomputer\cite{54}.

Large-scale gene identification, annotation and clustering
expressed sequence tags are another large scale problem in
genomics. Furthermore, it is well known that multiple genome
comparisons are essential and will constitute a significant
challenge in computational biomedicine. Understanding of human
diseases relies heavily on figuring out the intracellular
components and the machinery formed by the components. With DNA
microarrays, gene expression profiles in cells can be mapped
experimentally. Collective analysis of large number of these
microarrays across time or across treatment involves significant
computational tasks.

Genes are known to translate into protein and become the workhorse
of cell. The mechanistic understanding of biochemistry of the cell
involves intimate knowledge of the structure of these proteins and
details of their function. The number of genes from various
species are in the millions and computational modeling and
prediction of protein called protein folding is regarded as the
holy grail of biochemistry. The IBM Blue Gene project
\cite{IBMBlue01} estimates that simulating 100 microseconds of
protein folding takes $10^{25}$ machine instructions. This
computation on a Petaflops system will take three years or keep a
$3.3GHz$ microprocessor busy for the next million centuries. The
problem remains computationally intractable with modern
supercomputers even when knowledge-based constraints are employed.
Computer simulations remains the only way to understand the
dynamics of macromolecules and their assemblies. The simulations
which scale as $\OO{N^2}$ where $N$ is the number of atoms, are
still not capable of calculating motions of hundreds of thousands
of atoms for biologically measurable time scales.

Understanding the characteristics of protein interaction networks
and protein complex networks is another computationally intensive
problem. These small-world networks fall into three categories:
topological, constraint-driven, and dynamic. Each of these
categories involves complex combinatorial, graph theoretic, and
differential equation solver algorithms and could challenge any
supercomputer. With the knowledge of genome and intracellular
circuitry, precise and targeted drug discovery is possible. This
emerging computational field is a preeminent challenge in
biomedicine. ~\cite{54,Bader04}

\subsection{Astronomy and Astrophysics}
Astronomy is the study of the universe as a whole and of its
component parts of past, present and future. Observation is
fundamental in astronomy and controlled experiments are extremely
rare. The evolutionary time scales for most astronomical systems
are so long that these systems seem frozen in time, thus
constructing an evolutionary system from observation is therefore
difficult. An evolutionary model is constructed  from observations
involving many different systems of the same type (e.g. stars or
galaxies) at different stages and putting them in a logical order.
A HPC evolutionary model ties together these different stages
using known physical laws and properties of matter. The physics
involved in stellar evolution theory is complex and nonlinear,
thus without HPC, it is difficult to make significant advances in
the field. HPC can be used to turn a two-dimensional simulation of
a supernova explosion into a three-dimensional simulation or add
new phenomena into a simulation~\cite{54}. Simulation is an
important tool for astrophysicists to address different problems
and questions about galaxy formation and interaction, star
formation, stellar evolution, stellar death, numerical relativity,
and data mining of astrophysical data. The storage requirement for
simulation grows to more than $1$ Petabytes and the memory
requirements is more than $10$ Terabytes. Computational methods
such as Fast Multipole Method (FMM), Multi-scale dense linear
algebra, Parallel $3$D FFTs, Spherical Transforms, Particle
Methods and Adaptive Mesh Refinement are extensively used for
simulations~\cite{Simon04}.

\subsection{Computational Material Science and Nanotechnology}
The field of computational material science examines the
fundamental behavior of matter at atomic to nanometer length
scales and picosecond to millisecond time scales in order to
discover novel properties of bulk matter for numerous important
practical uses. Major research efforts include studies of:
electronics, photonics, magnetics, optical and mechanical
characteristics of matter; transport properties, phase
transformations, defect behavior and superconductivity in
materials and radiation interactions with atoms and solids.
Predictive equations that take the form of first principles
electronic structure molecular dynamics (FPMD)~\cite{Car97} and
Quantum Monte Carlo (QMC) are used for simulation of
nanomaterials. The computational requirement for this field grows
in the range of $\OO{N^3}$--$\OO{N^7}$, where $N$ is the number of
atoms in any simulations, making it an unlimited consumer of
increases in computing power. A practical application requires
large numbers of atoms and long time scales, in excess of what is
possible today. Revolutionary materials and processes from
material science will require petaflops of computing power soon.
~\cite{54} Other computational algorithms used for simulation
include Quantum Molecular Dynamics (QMD), Dense Linear Algebra,
Parallel $3$D FFT and Iterative Eigen Solvers ~\cite{Simon04}.

\subsection{Computational Fluid Dynamics (CFD)}

CFD\cite{Kutler89,Bailey92} is concerned with solving problems
involving combustion, heat transfer, turbulence, and complex
geometries such as magnetohydrodynamics and plasma dynamics.
Models used in CFD are growing in size, complexity and detail for
higher accuracy in prediction, thus requiring more powerful
supercomputing systems. These problems exhibit a variety of
complex behaviors such as advective and diffusive transport,
complex constitutive properties, discontinuities and other
singularities, multicomponent and multiphase behaviors, and
coupling to electromagnetic fields. These problems are represented
as nonlinear Partial Differential Equations (PDEs) that are time
dependent, and of physical space variables (up to three variables)
or phase space (up to six variables). Some applications require as
much as $1$ Terabyte of disk space to store information generated
for visualization ~\cite{Rosario94}. For many organizations, CFD
is critical to accelerate product time-to-market and overall
efficiency, as engineering and product development departments aim
to meet design deadlines. Aerospace organizations depend on CFD to
predict performance of their space vehicles in different
environments. CFD has become an integral component in the design
and test process, and simulation of the motion of fluid within or
around launch vehicles. Before costly physical prototyping begins,
design engineers leverage on CFD to visualize designs to predict
how rockets and satellites will perform. By computationally
analyzing design variations ahead of physical testing, optimal
design efficiency can be reached at reduced cost. CFD revolves
around extensive use of numerical methods to solve PDEs. In order
to arrive at a realistic solution, higher grid resolution must be
used and solving it in a reasonable amount of time requires a huge
amount of computational power. Computational methods usually used
for simulation includes Finite Difference, Spectral, Finite
Volume, Pseudospectral and Finite Element Methods.

\subsection{Computational Physics}
A mathematical theory describing precisely how a system will
behave is often impossible to be solved analytically. Hence the
implementation of numerical algorithms to solve such problems are
necessary, where higher resolution grid for spatial and temporal
dimension gives better accuracy. The most challenging problem in
computational physics at the moment is from plasma
physics~\footnote{http://www.ofes.fusion.doe.gov/FusionDocs.html}.
The main goal in plasma physics research is to produce
cost-effective, clean, and safe electric power from nuclear
fusion. Very large simulation of the reactions has to be run in
advance before building the generating device, thus saving
billions of dollars. Fusion energy, the power source of the sun
and other stars, occurs when the lightest atom, hydrogen, combine
to make helium in a very hot ($\approx100$ million degrees
centigrade) ionized gas, or ``plasma''. This field is a
computational grand challenge because, in addition to dealing with
space and time scales that can span more than 10 orders of
magnitude, the fusion-relevant problem involves extreme
anisotropy; the interaction between large-scale fluid-like
(macroscopic) physics and fine-scale kinetic (microscopic)
physics;and the need to account for geometric detail. Furthermore,
the requirement for causality (inability to parallelize over time)
makes this problem among the most challenging in computational
physics~\cite{54}. Computational methods usually used in plasma
physics are Gyrokinetic (GK), Gyro-Landau-fluid (GLF), nonlinear
solvers, adaptive mesh refinement, dense linear algebra and
particle methods~\cite{Ahearne04,Simon04}.

\subsection{Geophysical Exploration and Geosciences}

Geoscience is the study of the Earth and its systems.
Geoscientists design and implement programs to identify, delineate
and develop oil and natural gas deposits and reservoirs, coal
deposits, oil sands and nuclear fuels and nuclear waste
repositories. Numerical simulation is an integral part of
geoscientific studies to optimize petroleum recovery. Differential
equations are used to model the flow in porous media in three
dimensions. The need for increased physics of compositional
modeling and the introduction of geostatically based geological
models increases the computational complexity. Scientific study of
the Earth's interior such as geodynamo (an understanding of how
the Earth's magnetic field is generated by magnetohydrodynamic
convection and turbulence) in its outer core is a grand challenge
problem in fluid dynamics. HPC also plays a major role in the
understanding of the dynamics of Earth's plate tectonics and
mantle convection. This study requires simulation to incorporate
multirheological behavior of rocks that results in a wide range of
length scales and time scales, into three dimensional, spherical
model of the entire Earth. Computational methods such as
continuous Galerkin Finite Element Methods or Cell-centered Finite
Differences, Mixed Finite Element, Finite Volume, and Mimetic
Finite Differences are used for these simulations
~\cite{Spissue05}.

\medskip
\subsection{Summary}
In this section, we studied a variety of grand challenge
applications, that make use of different fundamental algorithms
and numerical methods. Each of these algorithms have different
computational, storage, memory and communication complexities.
Embarrassingly parallel, data parallel and parametric
problems that do not require significant communication can be
efficiently parallelized but problems that require significant
communication put a limit to achievable speedup. As the size of
the problem grows, the use of computational resources that are
geographically distributed is inevitable. This approach of
computing introduces many challenges due to the inherent dynamism
in computing resources and the Internet. Computational models
come into play here to provide a guideline of expected
performance available for a particular application,
as the application and given architecture continue to scale up.

In the next section, we look at a variety of HPC architectures
used to solve some of the computationally intensive applications
that we surveyed in this section.

\section{HPC Architectures}\label{sec3}

The first supercomputers called \emph{IBM $7030$ Stretch} and
\emph{UNIVAC LARC Sperry Rand } were functional in the early $1960$s.
In later years, supercomputers such as IBM $360$ models which
incorporate multiprogramming, memory protection, generalized
interrupts, $8$-bit byte, instruction pipelining, prefetch and
decoding, and memory interleaving were used. The U.S. supercomputer
industry was dominated by two companies: CDC and Cray Research.
Seymour Cray, better known as the father of supercomputers was
working with CDC in his earlier stage of his career, before he
founded Cray Research. These two companies are the only ones that
dominated the global supercomputer industry in the $1970$s and most of
$1980$s. During this period, Japan has also ventured into
the supercomputing industry two years after the first successful
commercial vector computer Cray-1 was shipped to them in $1976$.
Japans first vector processor known as FACOM $230$-$75$ APU (Array
Processing Unit) was installed at the National Aerospace
Laboratory in $1978$ ~\cite{52}. A few decades later the computing
technology has grown exponentially such that desktop computers have
become much more powerful than supercomputers in $1970$s and $1980$s.

It is anticipated that a petaflops capable supercomputer to be
available by $2008$.~\cite{Feldman06} At the time of writing,
Riken, (a Japanese government funded science and technology
research organization) has developed a supercomputer that achieves
a theoretical peak performance of one petaflops. However, the
system was not tested using Linpack so no direct comparison with
other benchmarked machines can be made.~\cite{editor06}
\reftable{evolution_of_SC} depicts the system parameters for the
fastest supercomputers built and used from $1997$ to $2006$. The
trend shows significant improvement in communication bandwidth for
both processor-memory and inter-processor communication, storage
capacity, and number of CPUs for more recent supercomputers.  Some
of the current (year $2004$ - $2006$) top high performance
computing architectures are listed in \reftable{recent_SC}. Note
that the cluster based architectures in some cases are
outperforming specialized supercomputer architectures based on the
rankings from the Top500 supercomputer list.

\begin{figure*}[htbp]
\centerline{
\psfrag{TERAGRID}{TeraGrid}
\includegraphics[width=4.5in]{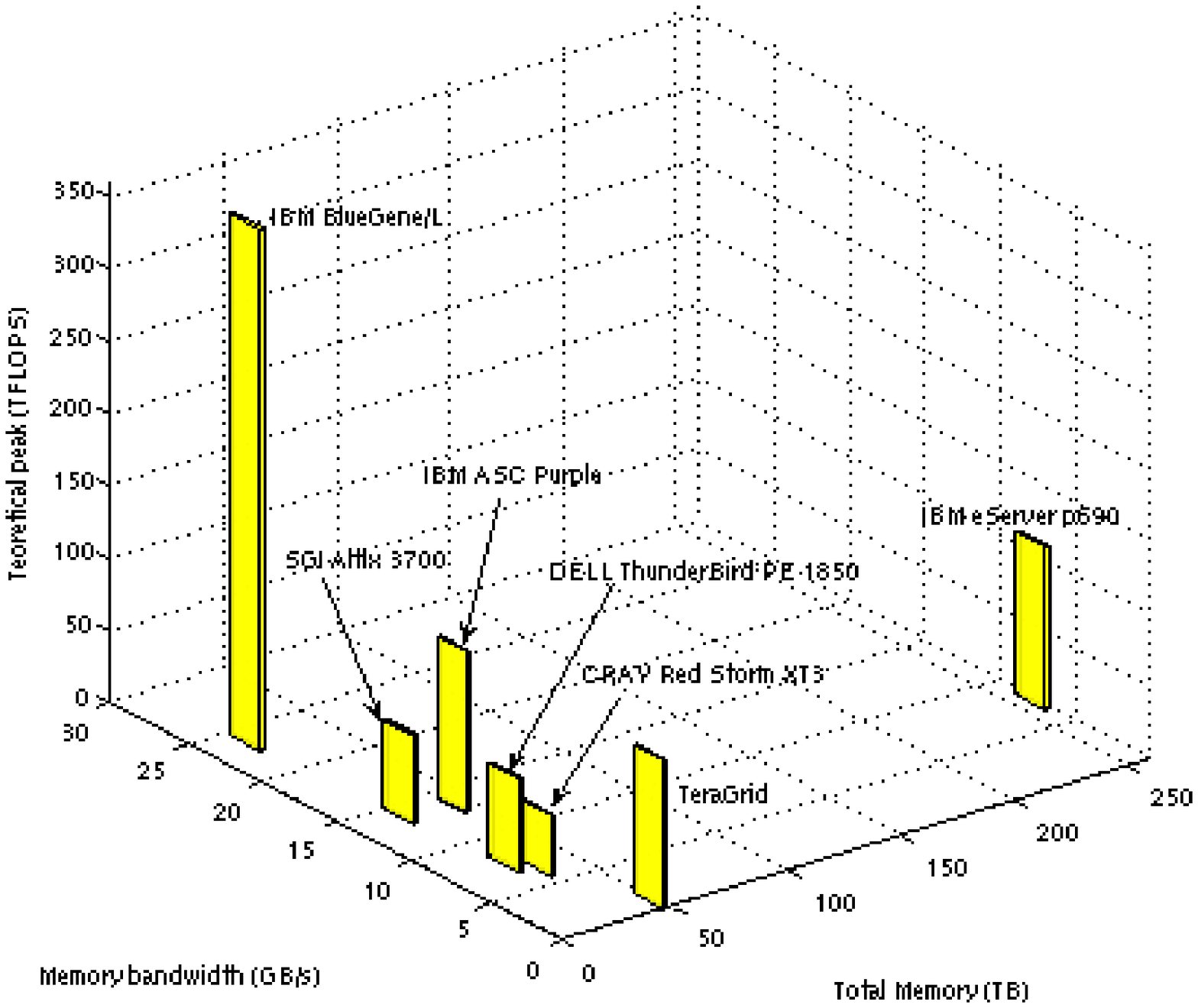}}
\caption{\label{sc_arch1} Theoretical peak, memory bandwidth and
total memory for some of the recent supercomputers.}
\end{figure*}

\begin{center}
\footnotesize{
\begin{longtable}{p{1in} p{1in} p{1in} p{1in} p{1in} }
\caption{\label{evolution_of_SC}System parameters for fastest
Supercomputers from $1997$ to $2006$. UKWN represents unknown
values.} \\
  \hline \\
  Model &IBM ASCI Red & IBM ASCI White & NEC Earth Simulator & IBM BlueGene/L \\
  \hline
  \endhead
  Fastest in Year & $1997-1999$ & $2000-2001$ & $2002-2003$ & $2004-2006$ \\
  Max. Memory (TB) & $1.212$ & $4$ & $10$ & $16$ \\
  LINPACK benchmark performance (TFLOPS) & $2.38$ & $7.304$ & $35.86$ & $280.6$ \\
  Max. \# Processors & $9632$ & $8192$ & $5120$ & $131072$ \\
  Clock cycle (GHz) & $0.2$ & $0.337$ & $0.5$ & $0.7$\\
  Memory B/W (GB/s) &$0.533$ & $2$ & $64$& $22.4$ \\
  Inter-node Comm. B/W (GB/s) & $0.8$& $0.5$& $12.3$ x $2$ & $3$D Torus:$0.175$, Tree network $0.35$ \\
  Operating system & TFLOPS OS & AIX & SUPER-UX &  CNK/LINUX \\
  Connection structure & $3$-D Mesh & $\Omega$-Switch & Multistage crossbar switch&  $3$-D Torus, Tree network, barrier network  \\
  Network interface & Network Interface Chip (NIC) and Mesh Interface Chip (MIC) & Ethernet,Token Ring, FDDI and other can be used & Crossbar switches & Gigabit Ethernet\\
  Cost & UKWN & UKWN & UKWN & $\geq$USD$1.5$M depending on configuration\\
  Applications  & Simulate the effects of massive nuclear explosions. & Stockpile Stewardship Program. &
    Earthquake, weather patterns and climate change including global warming.  & Scientific simulation and Stockpile Stewardship Program, Biomolecular simulation, computational fluid dynamics and molecular dynamics. \\
  Storage Capacity (TB) & $12.5$ & $160$ & $640$ & $400$ \\
  Processor type & IBM RS/$6000$ SP.& SP Power$3$ $375$ MHz& $8$-way replicated vector processor.  & PowerPC $440$  \\
  \hline
\end{longtable}
}
\end{center}

\begin{figure*}[htbp]
\centerline{\includegraphics[width=4.5in]{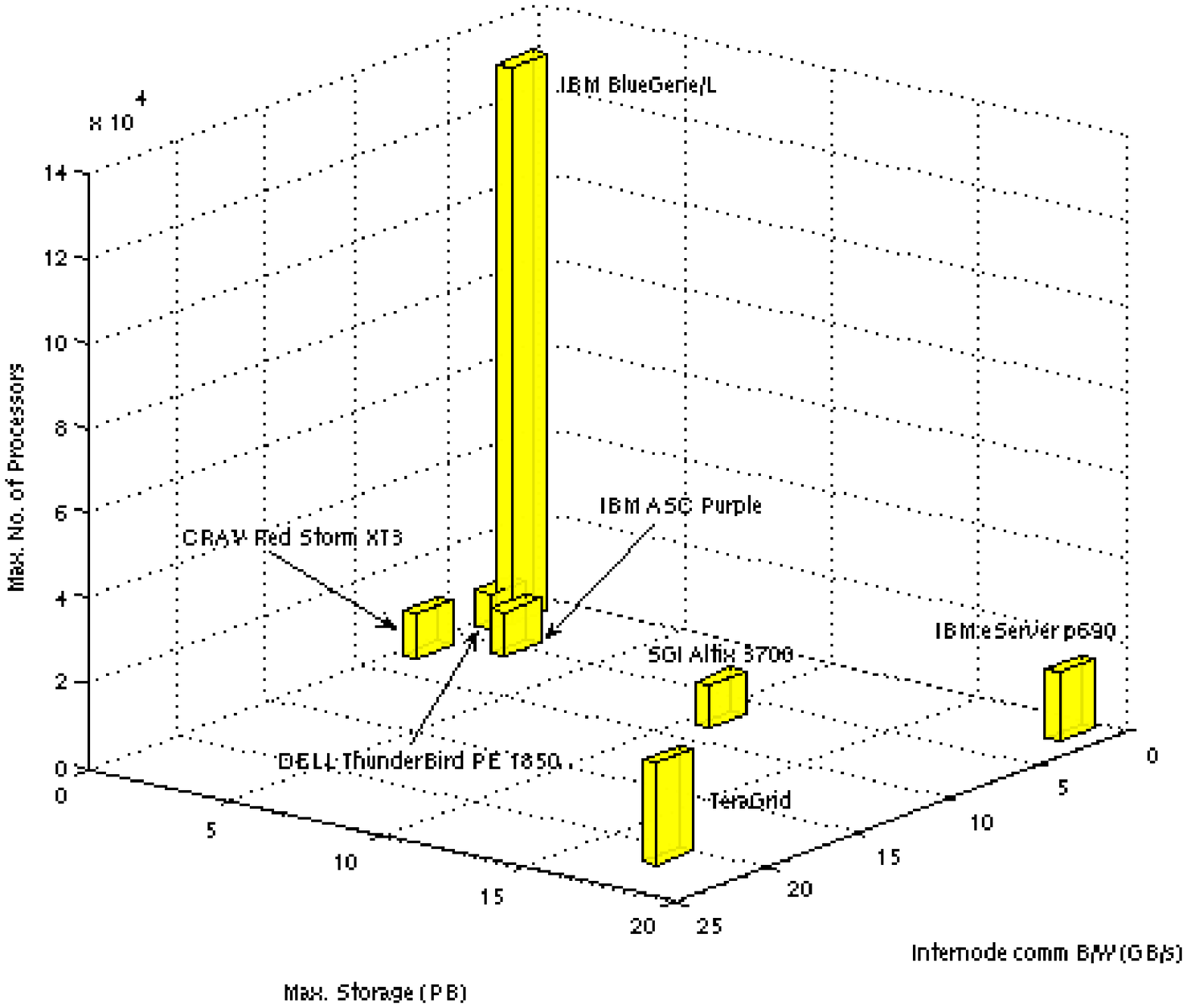}}
\caption{\label{sc_arch2} Internode communication bandwidth,
maximum number of processors and maximum storage available for
some of the recent supercomputers.}
\end{figure*}

\begin{center}
\footnotesize{
\begin{longtable}{p{1.0in} p{0.8in} p{0.8in} p{0.8in} p{0.8in} p{0.8in} p{0.8in}}
\caption{\label{recent_SC}Characteristic of some recent fast HPC
architecture. UKWN signifies an unknown entity and N/A stands for
Not Applicable.} \\
  \hline\\
   Vendor & IBM  & CRAY & DELL & SGI & IBM  & TeraGrid  \\
  \hline
  \endhead
  Model & BlueGene/L & Red Storm Cray XT$3$ & Thunderbird - PowerEdge $1850$  & NASA Columbia ALTIX $3700$ & ASC Purple & TeraGrid\\
  Available Memory(TB) & $16$ & $31.2$ & $24$ & $20$ & $40.96$ & $>45$ \\
  Cache &$32$KB L1; $2$KB L2; $4$MB L3 &$128$KB L1; $1$MB L2 & $2$MB L2& $32$KB L1; $256$KB L2; $6$MB L3 & $96$KB L1; $1.9$MB L2; $36$MB L3 & N/A \\
  Dist. Memory Architecture & Yes & Yes & Yes & No & Yes & Yes \\
  Architecture Type & MPP & MPP & Cluster & MPP & MPP & Grid \\
  Theoretical Peak (TFLOPS) & $360$ & $41.47$ & $64.512$ & $60.96$ &$111$ & $>102$\\
  Year (Ranking in Top500 list) & $2004$($\#1$), $2005$($\#1$)& $2005$($\#6$)& $2005$($\#5$) & $2005$($\#4$) & $2005$($\#3$) & $2006$(N/A)\\
  Max. \# processor & $131072$ &$10368$& $8192$ & $10240$ &$10240$ & $>24000$\\
  Operating system & Linux & Linux/Catamount & Linux & Linux & AIX & Heterogeneous\\
  Connection structure & $3$-D Torus, Tree Network & $3$-D Mesh $(27$x$16$x$24)$ & Classified (Red) and Unclassified (Black) &
  Crossbar and hypercube & Bi-directional, Omega-based variety of Multistage Interconnect Network (MIN) & Heterogeneous (Myrinet, SGI NUMAlink,
  InfiniBand, IBM Federation, $3$-D torus, global tree, Quadrics, Cray Seastar, Gigabit Ethernet and Sun Fire Link) \\
  Interconnect & Gigabit Ethernet & $100$ MB Ethernet& Infiniband & SGI Numalink, InfiniBand network, Gigabit Ethernet & Federation & Hub: CHI, ATL, LA, DEN, Abilene. (for connection between sites) \\
  Memory bandwidth (GB/s) & $22.4$ & $5.304$& $6.4$ & $12.8$ & $12.4$ & N/A\\
  Internode Comm. bandwidth (GB/s)& $\leq1.05$ & $6$ & $1.8$ & $6.4$ & $4$ & $10$-$30$ to Hub\\
  Cost & $\geq$USD$1.5$ depending on configuration & UKWN & UKWN & UKWN & UKWN & N/A\\
  Application specific & No & Yes & No & No & No & No\\
  Storage (PB) & $0.4$ &$0.24$ & $0.17$ & Online: $0.44$ Fibre channel RAID; Archive: $10$ & $2$ & Online:$3$; Mass:$>17$\\
  Processor &  PowerPC $440$ & AMD x86--64 Opteron & Dual Intel Xeon EM64T & Intel IA-$64$ Itanium $2$&Power5 & $8$ distinct architectures \\
  Clock speed (GHz)/processor & $0.7$ & $2.0$ & $3.6$ & $1.5$ &$1.9$ & N/A\\
  Site & DOE/NNSA/ LLNL & Sandia National Laboratories & Sandia National Laboratories & NASA/Ames Research Center/NAS &
  Lawrence Livermore Computing& ANL/UC/IU/ NCSA/ORNL/ PSC/Purdue/ SDSC/TACC\\
  \hline
\end{longtable}
}
\end{center}

In this section, we look at some of the HPC architectures that
consists of MPP, Cluster and Grids. \reffig{sc_arch1} and
\reffig{sc_arch2} shows the characteristics for some of the
supercomputers. It is interesting to note that the number of
processor used in recent architectures are increasing and hence
the increase in the peak performance. However, this peak
performance is not usually achievable because of other overheads
such as communication between nodes and data access from external
storage. The sustained performance of an architecture very much
depends on the type of application that is run, which relies on
algorithms, computational and communication complexity, size
of data that needs to be processed or generated for visualization
purposes. In general, to obtain more processing power, new
architectures are using more processors with higher memory
bandwidths compared to their predecessors. They also tend to have
large main memory and storage space to solve large scale problems
that incorporates high degree of abstraction and resolution size
for better accuracy. In the following sections we look at some of
the recent supercomputer characteristics in detail.

\subsection{IBM (Blue Gene/L)}
Blue Gene/L~\cite{Gara05,Adiga05,Ohmacht05} compute chip is a dual
processor (clock speed per processor $0.7$ GHz) system-on-a-chip
capable of delivering an arithmetic peak performance of $5.6$
Gigaflops. It is a Massively Parallel Processor (MPP) with
three-level on-chip cache that offers high-bandwidth and
integrated prefetching cache hierarchy on L2 ($32$ KB), L3 ($4$
KB) to reduce memory access time. Memory to CPU bandwidth of
$22.4$ GB/s is provided to serve speculative pre-fetching demands
of two processors cores~\cite{Ohmacht05}. The Blue Gene can be
scaled up to $65,536$ compute nodes yielding a theoretical peak of
$367$ Teraflops and has storage space of $400$ Terabytes
~\footnote{http://www-03.ibm.com/servers/deepcomputing/pdf/bluegenesolutionbrief.pdf}.
The nodes are interconnected through five networks: 1) a
$3$-dimensional torus network for point-to-point messaging between
computing nodes with a bandwidth of $0.175$ GB/s. If all six
bidirectional links that connect to a given node are fully
utilized, a bandwidth up to $1.05$ GB/s can be achieved; 2) a
global collective network for collective operation over the entire
application; 3) a global barrier and interrupt network; 4) a
gigabit Ethernet for machine control; and 5) another gigabit
Ethernet network for connection to other systems~\cite{Adiga05}.

\subsection{CRAY (Red Storm XT3)}
Red Storm is a MPP supercomputer at Sandia National Laboratories,
New Mexico. Red Storm was uniquely designed by Sandia and Cray,
Inc. It runs on $10,368$ AMD Opteron microprocessor at a clock
speed of $2$ GHz with a total memory of $31.2$ TB. Together with a
two level-on-chip cache memory hierarchy, $128$ KB L1 and $1$ MB
L2, and yields a theoretical peak of $41.47$ Teraflops. The system
provides a maximum of $5.304$ GB/s data flow between the cpu and
memory. It is constructed from commercial off-the-shelf parts
supporting IBM-manufactured SeaStar interconnect chip. The
interconnect chips, accompanies each of $10,368$ compute node
processors and is a key to three-dimensional mesh that allows
$3$-D representation of complex problems. The system has $6$ GB/s
CPU memory bandwidth and a storage space of $240$
Terabytes.~\footnote{http://www.cray.com/products/programs/red\_storm/index.html}
This architecture was built specifically for running simulation
for nuclear stockpile work, weapons engineering and weapons
physics.

\subsection{Dell Thunderbird }
ThunderBird~\footnote{http://www.cs.sandia.gov/platforms/Thunderbird.html}
is a supercomputer with cluster architecture at Sandia National
Laboratory running on a single core SMP node with dual Intel Xeon
EM64T processors. A total of $8,192$ processor at clock speed of
$3.6$ GHz is used. ThunderBird has a $2$ MB L2 cache memory and
$24$ Terabytes of main memory. With CPU memory bandwidth of $6.4$
GB/s it yields a theoretical speed of $64.5$ Teraflops.
Thunderbird has an interprocessor communication bandwidth of $1.8$
GB/s over $4$ InfiniBand network and a storage space of $170$
Terabytes~\cite{Singer05}.

\subsection{SGI (NASA Columbia ALTIX 3700)}
NASA's Columbia supercomputer is a MPP architecture with $10,240$
processor system comprising of twenty $512$-processor nodes.
Twelve of which are SGI Altix $3700$ nodes, and the other eight
are SGI Altix $3700$ Bx2 nodes. Each node is a shared memory,
Single System Image (SSI) system, running a Linux based operating
system. Four of the Bx2 nodes are linked to form a $2,048$
processor shared memory environment. It is powered by Intel IA-64
Itanium processor running at clock speed of $1.5$ GHz. it has
three-level on-chip cache of $32$ KB L1, $256$ KB L2 and $6$ MB L3
with CPU memory bandwidth of $12.8$ GB/s. The system has a maximum
theoretical peak of $60.96$ Teraflops. All the nodes are
interconnected via SGI Numalink, InfiniBand network and gigabit
ethernet network. It has an internode communication bandwidth of
$6.4$ GB/s and a combined storage space of $10.44$ Petabytes.

\subsection{IBM (ASC Purple)}
Each IBM ASC
Purple~\footnote{http://www.llnl.gov/computing/tutorials/purple/index.html}
node is a Symmetric multiprocessor (SMP)  powered by $8$ Power5
microprocessor running at $1.9$ GHz, configured with $32$ GB of
memory. The system at Lawrence Livermore Computing Laboratory has
a total of $1,280$ nodes with a combined total memory of $40.96$
TB. It has three-level-on-chip cache memory, $96$ KB L1, $1.9$ MB
L2, and $36$ MB L3 to reduce memory access time. A  CPU memory
bandwidth of $12.4$ GB/s comes together with a total number of
$10,240$ processors, so the theoretical speed achievable by this
system is $111$ Teraflops. The system also has a storage space of
$2$ Petabytes. All of the $1,280$ nodes in IBM ASC Purple system
are interconnected by dual plane federation (pSeries High
Performance) switch~\cite{Vendor05}. The federation network can be
classified as bidirectional, $\Omega-$based variety of Multistage
Interconnect Network (MIN). Bidirectional here refers to each
point-to-point connection between nodes comprised of two channels
(full duplex) that can carry data in opposite directions
simultaneously. MIN is used as an additional intermediate switch
to scale the system upwards.

\subsection{TeraGrid}
TeraGrid~\footnote{http://www.teragrid.org/}$^,$\footnote{http://www.teragrid.org/userinfo/hardware/index.php}
is an open scientific discovery infrastructure combining resources
at nine partner sites to create an integrated, persistent
computational resource. The partner sites are University of
Chicago, Indiana University, Oak Ridge National Laboratory,
National Center for Supercomputing Applications, Pittsburgh
Supercomputing Center, Purdue University, San Diego Supercomputer
Center, Texas Advanced Computing Center, and University of
Chicago/Argonne National Laboratory. TeraGrid integrates data
resources and tools, and high-end experimental facilities at all
the partners' sites using high-performance network connections.
These integrated resources have a combined $102$ Teraflops of
computing capability and more than $15$ Petabytes of online and
archival data storage with rapid access and retrieval over
high-performance networks. Researchers can access over 100
discipline-specific databases through TeraGrid. With this
combination of resources, TeraGrid is the world's largest
distributed infrastructure for open scientific research.

\subsection{Summary}
In this section, we looked at some of the recent supercomputers
and their characteristics. New supercomputers typically consume
less energy with higher computing capability. For example, NEC
Earth Simulator consumes $12,000$ kW power~\cite{Cameron05}
compared to $1,800$ kW power~\cite{Feng05,Gara05} by BlueGene/L
each producing $35.86$ TeraFlops and $280.6$ TeraFlops
respectively on LINPACK benchmark. Current HPC architectures have
higher memory bandwidth, a large number of processors and large storage
capacity compared to their previous generations. The current
fastest supercomputer, IBM BlueGene/L, was built to provide cost
effective performance but is not meant for all
applications~\cite{Gara05}. Here, a suitable parallel computing
model can be used to determine how an application can be
efficiently implemented on a given architecture. More importantly,
performance of a given architecture depends on the
configuration of the architecture and also the type of algorithm
that is used.

It is also worth noting that aggregating HPC resources
distributed across the WAN is becoming a trend in HPC as
demonstrated by the TeraGrid infrastructure. This is in part
contributed by the network technologies that are advancing at a
faster rate now compared to a decade ago. The power of network,
storage and computing resources are projected to double every $9$,
$12$ and $18$ months, respectively. Improvements in wide area
networking makes it possible to aggregate distributed resources in
collaborating institutions to solve problems in the area of
scientific computing using numerical simulation and data analysis
techniques to investigate increasingly large and complex
problems~\cite{Casanova02}. 

In the following section, we cover different parallel computing
models that are used to develop high performance software that
solve computationally intensive problems on HPC architectures
efficiently.

\section{Computational models}\label{Computational_models}
\subsection{Background on models}\label{Background_model} It
is important to have a clear picture of the problems and
architectures in order to see the connection with the associated
computational models and to see how the models have and can be
evolved. In the previous two sections, we covered a variety of HPC
challenge problems and described a number of HPC architectures
that have been developed to address these challenges. In this
section, we cover the development of computational models that
connect the high-level problem solving environments and approaches
to the lower-level architectural characteristics. We also see that
computational models tend to put emphasis on the architectural
parameters.
\begin{figure*} \centerline{
{\includegraphics[height=1.2in,width=4.2in]{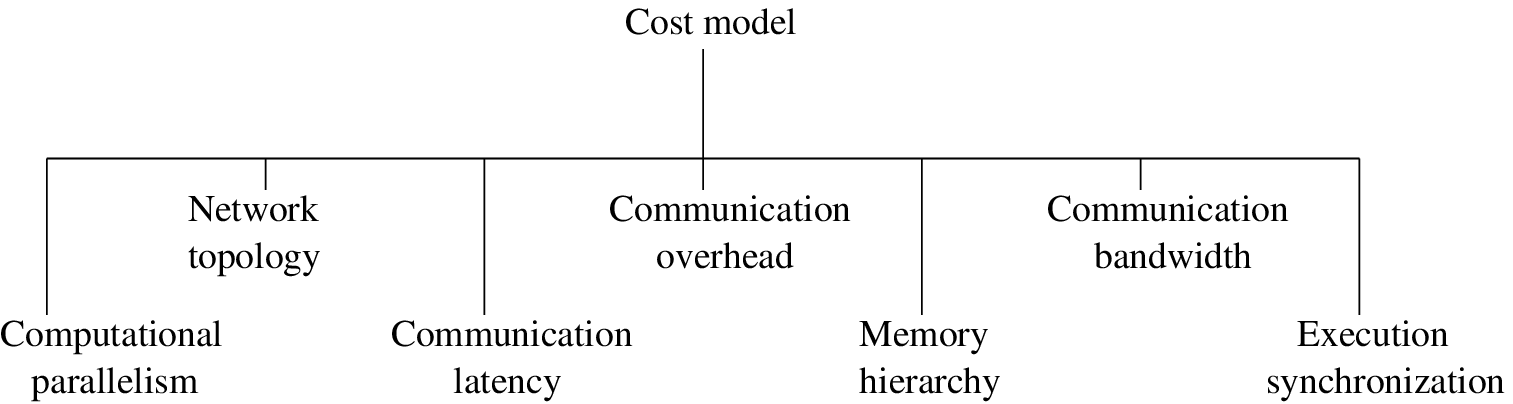}}}
\caption{\label{mcm}Characteristics of parallel architectures that
are emphasized in many traditional parallel computing models.}
\end{figure*}
It is common knowledge that a solution to any task begins with an
algorithm, which realizes the computational solution. However,
translating a problem to a computational algorithm requires a
model of computation that defines an execution engine. Thus, a
computational model plays an important role as a bridge between
software and hardware.

A model is said to be more powerful than another if algorithms
have a lower complexity in general on the machine. A computational
model also guides in the high-level design of parallel algorithms.
Models should balance between simplicity with accuracy,
abstraction with practicality, and descriptivity with
prescriptivity~\cite{6}. Models of parallel computation exists in
several levels. They are classified as: specification models (e.g.
Z~\footnote{The world Wide Web Virtual Library: The Z
notation,http://vl.zuser.org/}, VDM~\footnote{VDM Information,
http://www.csr.ncl.ac.uk/vdm/}, and CSP~\footnote{Virtual Library
formal methods;CSP, http://vl.fmnet.info/csp/}); programming
models (e.g. HPF~\footnote{HPF:The High Performance Fortran Home
Page, http://dacnet.rice.edu/Depts/CRPC/HPFF/index.cfm},
Split-C~\footnote{SPLIT-C,
http://www.cs.berkeley.edu/projects/parallel/castle/split-c/}, and
Occam~\footnote{OCCAM,
http://www.eg.bucknell.edu/~cs366/occam.pdf}); cost models (e.g.
PRAM~\cite{8}, BSP~\cite{4}, and LogP~\cite{13}); architecture
models (e.g. message-passing, RPC, shared memory, semaphores,
SPMD, MPMD) and physical models (e.g. distributed memory, shared
memory, and cluster of workstations and Grid). Despite the well
defined boundaries, there is some overlap by models: some
specifications act as programming models; some cost models act as
architectural models, etc~\cite{7}. In this section, we limit our
discussion domain on the cost model for accurate prediction of
parallel algorithm performance.

Many models have been developed for parallel architectures. The
majority of these models emphasize on seven important architecture
characteristics in parallel computing as depicted on \reffig{mcm}.
~\cite{6} These are:

\begin{description}
    \item[Computational parallelism] The number of processors,
    $p$, to be used in computation.
    \item[Network topology] Describes the inter-connectivity of
    processing nodes. Communication requirement of a parallel
    application should consider network topology of an
    architecture for efficient implementation.
    \item[Communication latency] Is the delay caused in accessing
    the non-local memory.
    \item[Communication overhead] Cost of message formation and
    injection of packets into the network.
    \item[Memory hierarchy] Is the different levels of memory from which data
    needs to be moved to reach the processor.
    \item[Communication bandwidth] Describes the bandwidth
    available for inter-processor communications.
    \item[Execution synchronization] The requirement for
    processors to wait until the required data has been received
    before proceeding with computations.
\end{description}

The Parallel Random Access Memory (PRAM) model was the most widely
used model~\cite{8}, with the assumption that all processors work
synchronously and communication between processor are costless. As
a result, the model has not been realistic in current parallel
architectures, where cost of communication delay, asynchrony and
memory hierarchy have far reaching impact on performance. These
constraints in the PRAM model provided sufficient catalyst to
develop models that emphasize on PRAM's weakness. Many variants of
the PRAM model have mushroomed ever since (e.g. Phase PRAM, APRAM,
LPRAM, and BPRAM). We will discuss them later in this section.
Other models that emphasize on weaknesses of the PRAM Model such
as the Postal model~\cite{24}, BSP (Bulk Synchronous Parallel)
~\cite{4} and LogP~\cite{13} considers communication costs such as
network latency and bandwidth. Parallel hierarchical models such
as Parallel Memory Hierarchy (PMH)~\cite{33}, Parallel
Hierarchical Memory model (P-HMM)~\cite{37}, LogP-HMM and
LogP-UMH~\cite{38} address the memory hierarchy in parallel
computing. \reftable{model_properties} shows some important
properties that are usually considered in parallel computing
models and the properties are explained below:

\begin{description}
    \item[Distributed/Shared memory] This property refers to type of memory used in a system that is
supported by the model. Shared memory system have multiple CPUs
all of which share the same address space. Whereas the distributed
memory system has in each CPU its own associated memory. The CPU
are connected by some form of network and exchanges data between
their respective memory when required.
    \item[Synchronous/Asynchronous] This property identifies if a model supports synchronous or
asynchronous algorithm.
    \item[Latency] Is the cost of accessing data in the memory (local,
shared or distributed memory). This property has significant
effect on performance of parallel algorithm. The cost increases
with the distance from the data requesting processor.
    \item[Bandwidth] Bandwidth in a HPC architecture can be divided into two parts
the memory and the interprocessor bandwidth. This bandwidth is not
unlimited and is an important characteristic to consider
particularly in distributed memory architecture.
    \item[Memory Hierarchy] This property denotes that the model takes into consideration
different level of memory hierarchy such as registers, cache, main
memory and secondary memory. This property is very important to
accurately reflect performance of an algorithm.
    \item[Overhead] Is the communication overhead introduced by processor for message
handling. It is defined to be the time the processor spends for
sending and receiving message. This value depends on the
communication protocol used.
    \item[Block transfer] This property takes into consideration the cost
of latency incurred when a block of memory is accessed. In most
architectures, cost of accessing the first address is expensive,
but accessing subsequent addresses is considerably cheaper.
    \item[Algorithms] List of algorithms that have been implemented or its
parallel complexity analyzed theoretically.
    \item[Architecture] Architectures used to analyze a particular model.
\end{description}

\subsubsection{Parallel Random Access Machine (PRAM) model and it's
variants}

The PRAM is an idealized parallel computing model that is widely
used to assess theoretical performance of parallel algorithms.
PRAM~\cite{8} is a shared memory model that has allowed
development of architecture independent parallel algorithms. Known
as an extension of RAM model, it mimics the processor part of RAM
model. A constant cost of memory access and computation steps are
assumed in this model. Since there maybe more than one
simultaneous memory read operation and simultaneous memory write
operation by processors, four different classes of PRAM model that
define how this should be handled is introduced~\cite{39}.

In the exclusive read, exclusive write (EREW PRAM) model, a memory
can only be accessed (for reading or writing) by one processor at
a time and it is the most restrictive model of the four. The
second model known as concurrent read, exclusive write (CREW
PRAM), allows a memory location to be accessed by more than one
processor simultaneously but only for reading the contents of the
locations. Memory access for writing can only be done one at a
time. The exclusive read, concurrent write (ERCW PRAM) model,
allows multiple processors to write but only one to read, this
model is usually not considered because a machine powerful enough
to support concurrent write should be able to accommodate
concurrent read. This model is thus subsumed in the CRCW model.
The fourth model, the concurrent read, concurrent write model
(CRCW PRAM), allows memory locations to be accessed by more than
one processor simultaneously for both reading and writing. For the
concurrent write permissable model (ERCW and CRCW) extra
specification is necessary to resolve how conflicts are overcome
and what the final stored result would be.

Absence of consideration for communication delay, asynchrony,
memory and network contention in PRAM has also contributed to its
lack of success. Consequently, many variations of the PRAM model
have been developed. The Phase PRAM~\cite{15} and APRAM~\cite{21}
model incorporates aspects such as asynchrony of processes. The
LPRAM~\cite{12} emphasizes on memory access. BPRAM (Block
PRAM)~\cite{16}, an extension of the LPRAM addresses communication
latency by considering the reduced cost for distributing a
contiguous block of data. Here we describe the purpose of the
variants and describe the functionality it plays in producing
better understanding in designing parallel algorithms and also in
predicting performance of parallel programs.

\begin{description}
\item[Phase Parallel Random Access Machine (Phase PRAM)] The Phase
PRAM~\cite{15} extends the PRAM model with partial asynchrony. Its
machine consists of a shared global memory, a set of $p$
sequential processors, and a local memory for each processor.
Computation is separated into a set of phases, and all processors
execute asynchronously, each phase is later ended by an explicit
synchronization. The cost of a synchronization step, $B(p)$, is
dependent on the number of processors $p$. This model discourages
too many inter-processor communication. Theoretical analysis and
simulation have been carried out for prefix sum, list ranking,
Fast Fourier Transform (FFT), bitonic merge, multiprefix, integer
sorting and Euler tours.~\cite{15}

\item[Asynchronous Parallel Random Access Machine (APRAM)] APRAM is
a ``fully" asynchronous model~\cite{21,22}. The APRAM model
consists of a global shared memory and a set of processes with
their own local memories. The basic operations executed by the
APRAM processes are called events. An APRAM computation is denoted
as the set of possible serializations of events executed by the
process. A virtual clock is associated with each serialization.
This virtual clock assigns a time $t(e)$ to each event $e$. The
clock "ticks" when each process has executed at least one event.
Events may be read and write events, which operate on the shared
and local memory, or local events. All events are charged unit
cost. The pair (round complexity, number of processes) is used to
measure the complexity of an APRAM algorithm, where a round is
defined as the sequence of events between two clock ticks in a
computation. The round complexity for a computation is defined to
be the maximum number of possible ticks for that computation. For
an algorithm the round complexity is defined as the maximum round
complexity over all of the possible computations~\cite{38}.
Complexity of graph connectivity and asynchronous summation
algorithms have been analyzed for this model.

\item[Local-Memory Parallel Random Access Machine (LPRAM)] The LPRAM
model~\cite{12} is a model that deals with bandwidth. It consists
of a shared global memory and a set of processors with unlimited
local private memory. The CREW PRAM is used to access global
memory and is more time consuming. At every time step, each
processor can perform either a communication step, in which it can
write and then read a word from the global memory, or a
computation step, which is an operation that accesses at most two
words from its local memory. Algorithms for matrix multiplication,
sorting and Fast Fourier Transform (FFT) have been implemented on
a binary tree architecture.

\item[Block Parallel Random Access Machine (BPRAM)] The BPRAM, which
is an extension of LPRAM~\cite{16}. BPRAM takes into consideration
the time saved in transmitting a contiguous block of data. The
model allows the usage of communication latency and the number of
processors and to determine the limits within which efficient
parallel algorithms can be written without taking into account the
details of the machine topology. Two parameters are used in the
BPRAM model, $l$ for startup cost or latency and $p$ the number of
processors, The cost of accessing local memory is taken in unit
time. For reading and writing a block size $b$ of contiguous
locations in global memory a cost of $l+b$ is charged. Theoretical
analysis for parallel algorithms such as matrix multiplication,
matrix transposition, rational permutation, permutation networks,
FFT and sorting have been investigated.

\end{description}

\begin{center}
\footnotesize{
\begin{longtable}{p{0.4in} p{0.5in} p{0.7in} p{0.2in} p{0.2in} p{0.3in} p{0.2in} p{0.3in} p{0.3in} p{1.8in}}
\caption{Properties incorporated in different models. In the
table, a check mark indicate that the characteristic is included
in the model.} \label{model_properties} \\
  \hline\\
  Models & \begin{sideways}Distributed or\end{sideways} \begin{sideways}Shared memory\end{sideways} & \begin{sideways}Synchronous\end{sideways} \begin{sideways}or Asynchronous\end{sideways} & \begin{sideways}Latency\end{sideways} & \begin{sideways}Bandwidth\end{sideways} &
  \begin{sideways}Memory\end{sideways}\begin{sideways}hierarchy\end{sideways} & \begin{sideways}Overhead\end{sideways} & \begin{sideways}Block\end{sideways} \begin{sideways}transfer\end{sideways}
  & \begin{sideways}Network\end{sideways} \begin{sideways}topology\end{sideways}& Architectures  \\
  \hline \\
  \endhead
  PRAM & Shared & Synchronous & & & & & & & Had been applied to many architectures but not accurate.
  \\ \\
  \multicolumn{10}{l}{Algorithms: Matrix multiplication, solving system of linear equation,
  sorting, FFT, Graph problems, etc.}\\
  \hline \\
   Phase PRAM & Shared & Semi-asynchronous & \ltcenter{\checkmark} & & & & & & \ltcenter{-}\\\\
  \multicolumn{10}{l}{Algorithms: Prefix sum, list ranking, FFT, bitonic merge, multiprefix,
  integer sorting and Euler tours. } \\
  \hline \\
  APRAM & Shared & Asynchronous & & & & & & & \ltcenter{-}\\\\
  \multicolumn{10}{l}{Algorithms: Graph connectivity and asynchronous summation.
  } \\
  \hline \\
  LPRAM  & Shared & Synchronous & &\ltcenter{\checkmark} & & & & & Binary tree. \\\\
  \multicolumn{10}{l}{Algorithms: Matrix multiplication, sorting and FFT.}\\
  \hline \\
  BPRAM & Shared & Synchronous & \ltcenter{\checkmark} & & & & \ltcenter{\checkmark} & & \ltcenter{-} \\\\
  \multicolumn{10}{l}{Algorithms: Matrix (multiplication, transposition),
  rational permutation, permutation networks, FFT and sorting.}\\
  \hline \\
  Postal model & Distributed  & Asynchronous & \ltcenter{\checkmark} & & & & & & \ltcenter{-} \\\\
  \multicolumn{10}{l}{Algorithms: Broadcast and summation.}\\
  \hline \\
  BSP & Distributed  & Semi-asynchronous & \ltcenter{\checkmark} & \ltcenter{\checkmark} &  & & & & Clusters, Network of workstations, multistage network etc. \\\\
  \multicolumn{10}{l}{Algorithms: NBody, Ocean Eddy, Minimum spanning tree (MST), Shortest path and Matrix multiplication.}\\
  \hline \\
  D-BSP & Both & & \ltcenter{\checkmark} & \ltcenter{\checkmark} & & & & & \ltcenter{-}\\\\
  \multicolumn{10}{l}{Algorithms: Sorting and routing.}\\
  \hline  \\
  E-BSP & Distributed & Semi-asynchronous & \ltcenter{\checkmark} & \ltcenter{\checkmark} & & & &\ltcenter{\checkmark}& Linear array and mesh network.\\\\
  \multicolumn{10}{l}{Algorithms: Matrix multiplication, routing problem, all-to-all broadcast and finite difference
  application.} \\
  \hline \\
  LogP & Both & Asynchronous & \ltcenter{\checkmark} & \ltcenter{\checkmark} & & \ltcenter{\checkmark} & & & Hypercube (nCUBE/2), Butterfly (Monsoon), Torus (Dash), $3$D mesh (J-Machine), Fat-tree (CM-5)\\\\
  \multicolumn{10}{l}{Algorithms: Parallel sorting, broadcast, summation, Fast Fourier Transform (FFT),
  and LU Decomposition.}\\
  \hline \\
  CGM & Both & Semi Asynchronous & &\ltcenter{\checkmark} & & & & & 2D Mesh, hypercube and fat-tree. \\\\
  \multicolumn{10}{l}{Algorithms: Geometric algorithms (e.g. $3$D-Maxima, multisearch on balanced search tree,
  }\\
  \multicolumn{10}{l}{$2$D-nearest neighbors of a point set etc.), Graph problems (List rankings,Euler tour construction,}\\
  \multicolumn{10}{l}{tree contraction and expression tree
  evaluation, etc.).}\\
  \hline \\
  PMH & Distributed & Asynchronous & \ltcenter{\checkmark} & \ltcenter{\checkmark} & \ltcenter{\checkmark} & \ltcenter{\checkmark}
  & \ltcenter{\checkmark}& &Tree, ring and 2-D Mesh. \\\\
  \multicolumn{10}{l}{Algorithms: }\\
  \hline \\
  P-HMM & Distributed & Asynchronous &  &  & \ltcenter{\checkmark} & & & & \ltcenter{-}\\\\
  \multicolumn{10}{l}{Algorithms: Matrix transpose and list ranking}\\
  \hline \\
  logP-HMM & Distributed & Asynchronous & \ltcenter{\checkmark} & \ltcenter{\checkmark} & \ltcenter{\checkmark} & \ltcenter{\checkmark}& & & Fat-tree (Thinking machine CM-5).\\\\
  \multicolumn{10}{l}{Algorithms: FFT and sorting }\\
  \hline \\
  logP-UMH & Distributed & Asynchronous & \ltcenter{\checkmark} & \ltcenter{\checkmark} & \ltcenter{\checkmark} & \ltcenter{\checkmark}& & & Fat-tree (Thinking machine CM-5).\\\\
  \multicolumn{10}{l}{Algorithms: FFT and sorting }\\
  \hline \\
  \end{longtable}
}
\end{center}
\subsubsection{Postal Model} The Postal model~\cite{24} is a
distributed memory model with the constraint that the
point-to-point communication has latency $\lambda$. It can be
regarded as a model described by two parameters:$p$ and $\lambda$,
where $p$ is the number of processors. Several elegant optimal
broadcast and summation algorithms have been designed based on
this model, which were then extended for LogP model~\cite{13}.
Algorithms other than broadcast and summation have largely not
been presented for this model.

\subsubsection{Bulk Synchronous Parallel (BSP) and it's variants}
BSP~\cite{4} model provides support for developing architecture
dependent model, thus indirectly promotes wide spread software
industry for parallel computing. It has a cost model which
incorporates essential characteristics of parallel machines. A BSP
program is one which proceeds in stages, known as
superstep.\footnote{http://users.Comlab.ox.ac.uk/bill.mccoll/oparl.html}
A superstep consists of computation, communication and
synchronization phases. In the first phase, processors compute
using locally held dataset. Data are then communicated between the
processors in the second phase. In the third phase,global
synchronization is carried out, and this is to ensure all the
messages involved in communication are received before moving on
to the next superstep. BSP parameters $p$, $g$, and $L$ are used
to evaluate performance of a BSP computer. $p$ represents number
of processor, $g$ and $L$ represents network parameters. If
maximum local computation in a step takes time $W$, and the
maximum number of send or receive by any processor is $h$ then the
total time for a superstep is given by $T=W+hg+L$. Algorithms for
N-Body, ocean Eddy, minimum spanning tree (MST), shortest path,
matrix multiplication, sorting and routing have been developed
using this model.~\cite{17,18,19,20}

\begin{description}
\item[LogP] The LogP model is motivated by current technological
trends in high performance computing towards networks of
large-grained sophisticated processors. The LogP model uses the
parameters $L$ for an upper bound of latency for transmitting a
single message, $o$ for computation overhead of handling message,
$g$ a lower bound of time interval between consecutive message
transmission at a processor and $P$ the number of
processors.~\cite{13}. In contrast to the BSP model, it removes
the barrier synchronization requirement (h-relation in BSP) and
allows the processors to run asynchronously. The network of a LogP
machine has a finite capacity such that at any time at most
$\lfloor L/g \rfloor$ messages can be in transit from or to any
processor. It can support both shared and distributed memory
architecture. The LogP model encourages well-known general
techniques of designing algorithms for distributed memory machines
including exploiting locality, reducing communication complexity,
and overlapping communication and computation. The LogP model also
promotes balanced communication patterns by introducing the
limitation on network capacity so that no processor is overloaded
with incoming messages. Moreover, it is often reasonable to ignore
parameter of $o$ in a practical machine, such as in a machine with
low bandwidth (high $g$). Parallel complexity analysis for
sorting, broadcast, summation, Fast Fourier transform (FFT) and LU
decomposition have been developed and implemented on different
architectures such as hypercube, butterfly, Torus, $3$D mesh, and
Fat-tree~\cite{karp93}.

\item[Coarse Grained Multi Computer (CGM)] CGM
\cite{26,27,28,25} is a version of BSP model, it allows only bulk
messages to be sent in order to minimize message overhead costs. A
CGM consists of a set of $P$ processors $P_1,P_2,\ldots,P_n$
processors. Each communication round consists of routing a single
$h-relation$ message. All information sent from one processor to
another processor is packed into one large message to reduce
communication overhead. Thus the communication time in CGM
computer is the same as BSP computer plus the packaging time. An
optimal algorithm in CGM model is equivalent to minimizing the
number of communication round as well as local computation time.
The model also minimizes other important costs such as message
overhead and synchronization time. Parallel complexity of
geometric algorithms (e.g. $3$D-Maxima, multisearch on balanced
search tree, $2$D-nearest neighbors of a point set etc.), graph
problems (List rankings, Euler tour construction, tree contraction
and expression tree evaluation) have been analyzed and implemented
on architecture such as $2$D Mesh, hypercube and fat-tree.

\item[Extended BSP (E-BSP)]
The BSP as well as BPRAM assume that the time needed for
communication is independent of the network load. The BSP model
conservatively assumes that all $h$-relations are full
$h$-relations in which all processors send and receive exactly h
messages. Likewise, in the BPRAM it is assumed that sending one
$m$-byte message between two processors takes the same amount of
time as a full block permutation in which all processors send and
receive a m-byte message. The E-BSP model\cite{23} extends the
basic BSP model to deal with unbalanced communication patterns,
i.e., communication patterns in which the processors send or
receive have different data size. Like BSP, the E-BSP model is
strongly motivated by various routing results. Furthermore, the
cost function supplied by E-BSP generally is a non-linear function
that strongly depends on the network topology. Several algorithms
that uses this model such as routing problem, all-to-all broadcast
operation, matrix multiplication and finite difference application
have been developed.

\item[D-BSP]
Decomposable Bulk Synchronous Parallel(D-BSP)\cite{30,29} is a
variant of BSP to capture some aspects in network proximity. A set
of n processor/memory pairs that can be partitioned as a
collection of clusters, where each cluster is independent of the
other and is characterized by its own bandwidth and latency
parameters. The partition of clusters can change dynamically
within a pre-specified set of legal partitions. The advantage is
that communication patterns where messages are confined within
small clusters have small cost. Thus the model is claimed to
represents realistic platforms unlike as in standard BSP. This
advantage translates into higher effectiveness and portability of
D-BSP over BSP.

\end{description}

\subsubsection{Memory hierarchy models} As technology in
electronics matures, different components of computer improves at
different rates. In particular, the rate of increase in processor
speed is far more rapid compared to the increase in bandwidth for
local memory. Memory hierarchy was introduced in computer
architecture to assist in keeping up with the memory request rate
from central processing unit. This allows, data to be accessed
from the fastest memory, such that the average time for fetching
data is reduced significantly. Each level of memory in the memory
hierarchy has its own costs and performance. Thus to reduce cost,
memory that are more expensive to build is used stringently. At
the lowest level, CPU registers and caches are built with the
fastest and most expensive memory. At a higher level, inexpensive
but slower disks are used for external mass
storage~\cite{Vitter01}. Models that do not reflect the usage of
memory hierarchy is most likely to be inaccurate, because of the
presence of registers, caches, main memory and disks. Programs
that are tuned to a particular architecture by considering memory
hierarchy can produce significant speed up, thus it is important
to write programs that takes memory hierarchy into consideration.
As a result, computational models to reflect performance of these
programs are established. Data movement to and from processors,
cache memory and main memory incur some cost depending on the
distance from the processing unit. In the RAM model, there is no
concept of memory hierarchy; each memory access is assumed to take
one unit of time. This model ``may'' be appropriate for small size
of problem that can fit into the main memory, but as mentioned
earlier registers, cache and disks can contribute to inaccuracy.
Many variant of hierarchical memory model has been introduced, in
this section we discuss some of the models.
\begin{description}
\item[Parallel Hierarchical Memory Model (P-HMM)]
The Hierarchical Memory model (HMM) introduced by Agrawal et. al
~\cite{36} charges a cost of $f(x)$ to access memory location $x$
instead of a constant time taken in the Random Access Machine
(RAM)~\cite{Aho74} model. In HMM the concept of block memory
transfer to utilize spatial locality in algorithms was not
introduced but the Hierarchical Memory Model with Block Transfer
(HMBT)~\cite{Aggarwal87} takes this factor into consideration. The
P-HMM model is also known as the parallel I/O
model~\cite{42,vitter90}. This model considers data that resides
in hardisk rather than just the main memory. For allowing parallel
data transfer, the P-HMM was introduced. It has $P$ separate
memories connected together at the base level of the hierarchy.
Each $P$ hierarchies can function independently, and communication
between hierarchies takes place at the base memory level. The $P$
base memory level locations are interconnected via a network and
the $P$ hierarchies can each function independently. This model
also assumes that the $P$ base memory levels are interconnected
via a network such as a hypercube or cube-connected.~\cite{42}

\item[Parallel Memory Hierarchy (PMH)]
The PMH model\cite{33} uses a single mechanism to model the costs
of inter-processor communication and memory hierarchy. A parallel
computer is modeled as a tree of memory modules with modules at
the leaves as processors. The leaf module performs computation
while other modules holds data. Data in a module is partitioned
into blocks and it is the basic unit of data transfer between a
child and its parent. Communication between two processor
resembles somewhat like a fat-tree model but differs by having
memory and messages made explicit. The model has four parameters
for each module $m$, the block-size $s_m$ (number of bytes per
block of $m$ ); the block-count $n_m$ (number of block that fits
in $m$); the child-count $c_m$ (number of children $m$ has);
transfer time $t_m$ (number of cycles it takes to transfer a block
between $m$ and its parent). Appropriate tree structure and
parameter values should be chosen confirming to the machines
communication capabilities and memory hierarchy.

\item[LogP-HMM]
This model consist of two parts: the network and the memory part.
The network part is captured by LogP model and the memory part by
the Hierarchical Memory Model (HMM) thus the name
LogP-HMM.~\cite{38} This model is defined much like a P-HMM model.
It consists of a set of asynchronously executing processors, each
with an unlimited memory. Local memory is organized as a sequence
of layers with increasing size, where size of memory block is $1$
and the size of layer $i$ is  $2^i$. The cost of accessing a
memory location at address $x$ is $log\text{ }x$. The processors
are connected by LogP network at level $0$. It also assumes that
the network has finite capacity such that at any time at most
$\lfloor\frac{L}{g}\rfloor$ messages can be in transit from or to
any processor.

\item[LogP-UMH]
The primary difference between LogP-UMH~\cite{38} and LogP-HMM is
that the former uses memory organized as in Uniform Memory
Hierarchy (UMH)~\cite{31}. The UMH model is an alternative model
for multilevel memories and is an instance of the more general
Memory Hierarchy (MH)~\cite{31} model. The MH model consists
several memory module levels and each module is characterized by
three parameters: $s_l$ (the number elements in a block), $n_l$
(the number of blocks), and $b_l$ (the time to move a block of
size $s_l$ from level $l$ to level $l+1$).
$UMH_{\alpha,\rho,f(l)}$ is a simplification of MH model that
defines the $l$th memory level $M(l)$ as $M(l)=\langle s_l, n_l,
b_l \rangle$ = $\langle
\rho^{l},\alpha\rho^{l},\rho^{l}f(l)\rangle$, where $\alpha$ and
$\rho$ are integer constants. That is, the $l$th memory level
consists of $\alpha\rho^{l}$ blocks, each of size $s(l)=\rho^{l}$,
and is connected to levels $l-1$ and $l+1$. Each block on level
$l$ can be randomly accessed as a unit and transferred to or from
level $l+1$ with a cost of $\rho^{l}f(l)$, where $f(l)$ is a well
behaved function for the level $l$ and is known as the transfer
cost function ($\frac{1}{f(l)}$ is the bandwidth).

\end{description}

\subsection{Models for Wide Area Network (WAN)}
Parallel applications are traditionally run on dedicated
supercomputers where resources are usually homogeneous, with
predictable network behavior and are usually allocated entirely
for a single application without contention from other
applications. Developing computational model for grid environment
is difficult due to heterogeneous computing resources,
heterogeneous network (bandwidth and latency), resource contention
from different application, reliability and availability issues.
However, attempts are already made to estimate the
behavior/performance of parallel application on this environment.
In this section we discuss some of the works.
\subsubsection{Heterogeneous Bulk Synchronous Parallel- k (HBSP$^k$)}
The k-Heterogeneous Bulk Synchronous Parallel~\cite{Williams00}
(HBSP$^k$) model is a generalization of the BSP model~\cite{4} of
parallel computation. This model is characterized by eleven
parameters as shown in \reftable{hbspk} which can be used to
accommodate different architectures. HBSP$^k$ is claimed to
provide sufficient information for developing parallel
applications on wide-range of architecture such as traditional
parallel architecture (supercomputers), heterogeneous clusters,
the internet and computational grids. Each of these system are
then grouped together based on their ability to communicate with
each other.

\begin{longtable}[htbp]{p{0.6in}p{4.5in}}
\caption{Parameters used in HBSP$^k$ model.}
\label{hbspk} \\
\hline \\
Parameters & Description \\
\hline \\
$M_{i,j}$ & a machine's identity, with $0\leq i \leq k$, $0\leq j\leq m_i$. \\
$m_i$ & number of HBSP$^k$ machines on level $i$. \\
$m_{i,j}$ & number of children of $M_{i.j}$. \\
$g$ & A bandwidth indicator that reflects the speed at which the fastest machine can inject packets into the network.\\
$r_{i,j}$ & The speed relative to the fastest machine for $M_{i,j}$ to inject packets into the network.\\
$L_{i,j}$ & overhead to perform a barrier synchronization of the machines in the subtree of $M_{i,j}$.\\
$c_{i,j}$ & fraction of the problem since that $M_{i,j}$ receives.\\
$h$ & size of a heterogeneous $h$-relation. \\
$h_{i,j}$ & largest number of packets sent or received by $M_{i,j}$ in a super$^i$-step.\\
$S_i$ & number of super$^i$-step. \\
$T_i(\lambda)$ & execution time of super$^i$-step.\\
\hline \\
\end{longtable}

HBSP$^k$ refers to a class of machines with at most $k$ levels of
communication. When $k=0$ it represents a single processor system,
for $k=1$ it represents class of machines which consists of at
most one communication network, as an example, a HBSP$^1$ machine
may include a single processor systems(i.e. HBSP$^0$), traditional
parallel machines, and heterogeneous workstation clusters. In
general, HBSP$^k$ systems include HBSP$^{k-1}$ computers as well
as machines composed of HBSP$^{k-1}$ computers and the
relationship of the machine classes is HBSP$^0\subset$ HBSP$^1$
$\cdots\subset$HBSP$^k$.

\begin{figure}[htbp]\centerline{
\psfrag{M0,0}{$M_{0,0}$}\psfrag{M0,1}{$M_{0,1}$}
\psfrag{M0,2}{$M_{0,2}$}\psfrag{M0,3}{$M_{0,3}$}\psfrag{M0,4}{$M_{0,4}$}
\psfrag{M0,5}{$M_{0,5}$}\psfrag{M1,0}{$M_{1,0}$}\psfrag{M1,1}{$M_{1,1}$}
\psfrag{M1,2}{$M_{1,2}$}\psfrag{M2,0}{$M_{2,0}$}\psfrag{HBSP2}{HBSP$^2$}
\psfrag{super0-steps}{super$^0$-steps}\psfrag{super1-steps}{super$^1$-steps}
\psfrag{super2-steps}{super$^2$-steps}\psfrag{k=2}{$k=2$}
\psfrag{(M1,0)}{($M_{1,0})$}\psfrag{(M1,2)}{($M_{1,2})$}
\psfrag{(M2,0 & M1,1)}{($M_{2,0}$ \& $M_{1,1}$)} \psfrag{An HBSP2
cluster}{An HBSP$^2$ cluster}\psfrag{Tree representation of HBSP2
cluster}{Tree representation of HBSP$^2$ cluster}
{\includegraphics[height=2.35in,width=6in]{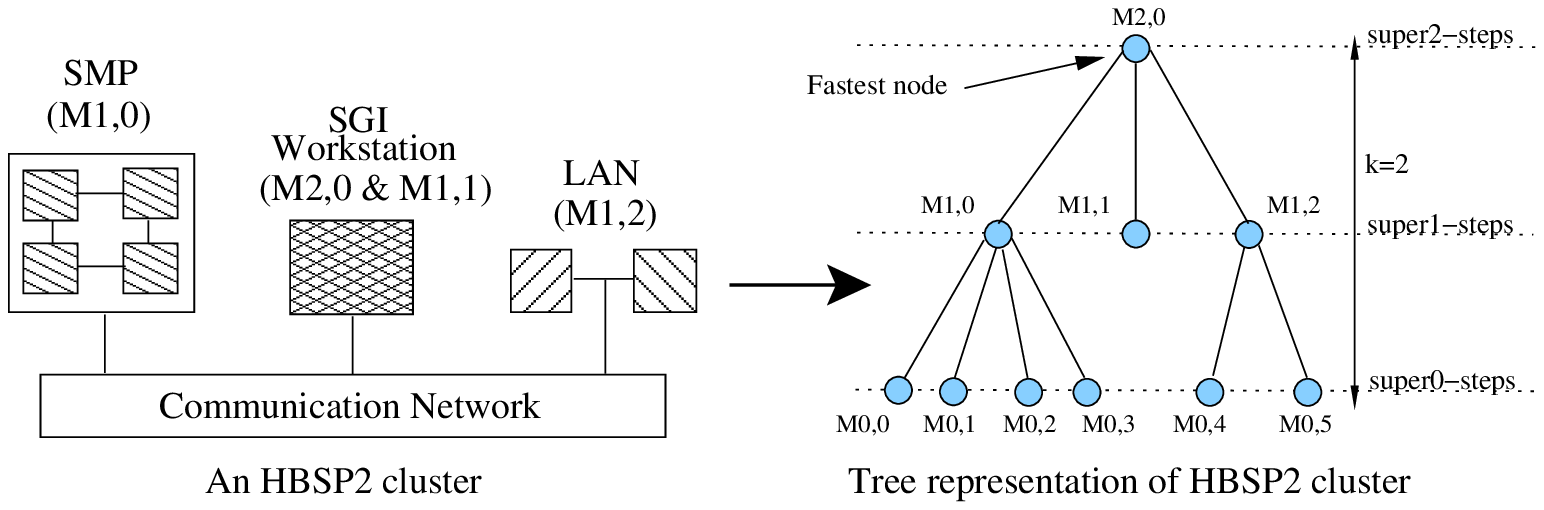}}}
\caption{An HBSP$^k$ cluster and it's tree representation. } 
\label{hbsp2}
\end{figure}

A HBSP$^k$ machine is represented by a tree $T=(V,E)$. Each node
of $T$ represents a heterogeneous machine. The level of root is
equal to the height of the tree, $k$ and root $r$ of tree $T$ is
known as a HBSP$^k$ machine. If $d$ is the length of the path from
the root $r$ to a node $x$, the level of node $x$ is $k-d$. Thus
nodes at level $i$ of tree $T$ are HBSP$^i$ machines.
\reffig{hbsp2} shows the HBSP$^2$ cluster and it's tree
representation in this model. Machines are indexed according to
level $i$, $0\leq i \leq k$, are labeled
$M_{i,0},M_{i,1},\ldots,M_{i,m_{i}-1}$, where $m_{i}$ represents
the number of HBSP$^{i}$ machines. Machine $M_{i,j}$ of a
HBSP$^{k}$ computer, where $0\leq j \leq m_{i,j}$ is a cluster
with identity $j$ on level $i$. A machine at level $i$ of tree $T$
is taken as a coordinator nodes of machines at level $i-1$. This
coordinators act as a representative for their cluster during
inter-cluster communication or represent the fastest computer in
their subtree to increase algorithmic performance. Cost of
computation by HBSP$^k$ machine is calculated directly at each
level $i$.

An HBSP$^k$ computation consists of a combination of
super$^i$-steps and during a super$^i$-step, each level $i$ node
performs asynchronously some combination of local computation,
message transmission to other level $i$ machines, and message
arrivals from its peers. A message that is sent in one
super$^i$-step is guaranteed to be available to the destination
machine at the beginning of the next super$^i$-step. This is
achieved by having a global synchronization of all the level $i$
computers after each super$^i$-step. A HBSP$^1$ machine has to
perform communication to transfer data, unlike HBSP$^0$ machine
where communication and synchronization is not applicable. A
HBSP$^1$ computation resembles a BSP computation but only differs
in how HBSP$^1$ algorithm delegates more work to the faster
processor. The HBSP$^2$ machine consists of super$^1$-steps and
super$^2$-steps. In the super$^2$-step, the coordinator nodes for
each HBSP$^1$ cluster performs local computation and/or
communication between other level $1$ coordinator nodes.

The value of $r_{i,j}$ for the fastest machine (root) is
normalized to $1$. Thus other machines, $M_{i,j}$, are said to be
$t$ times slower than the fastest machine if $r_{i,j}=t$. The
$c_{i,j}$ parameter is used for load balancing purposes, it
provides problem size to machine $M_{i,j}$ that is proportional to
its computational and communication capabilities. The HBSP$^k$
model does not mention about how to find values of $c_{i,j}$, and
assumes that the cost have been determined beforehand.

The execution time of super$^i$-step is given by,
\begin{equation}
T_i(\lambda)=w_i+gh+L_{i,j}. \nonumber
\end{equation}
where, $w_i$, represents the largest amount of local computation
performed by an HBSP$^i$ machine, $h$=max\{$r_{i,j}\cdot
h_{i,j}$\}, is the heterogeneous $h$-relation with $h_{i,j}$ the
largest number of messages sent or received by $M_{i,j}$, where
$0\leq j< m_{i}$ and $gh$ as the routing cost. If $S_i$ is the
number of super$^i$-steps, where $1\leq i \leq k$. The execution
time of an HBSP$^k$ algorithm is the total time taken by
super$^i$-steps. Thus the overall cost given by this model is,
\begin{equation}
\sum_{\lambda=1}^{S_1}T_1(\lambda)+\sum_{\lambda=1}^{S_2}T_2(\lambda)+\ldots+\sum_{\lambda=1}^{S_k}T_k(\lambda).
\nonumber
\end{equation}
This model shows factors that are important to be considered when
designing HBSP$^k$ application. Similar to BSP model, to minimize
the execution time, programmer must consider, $(i)$ balancing the
local computation of the HBSP$^k$ machines in each super$^i$-step,
$(ii)$ balance the communication between the machines, and $(iii)$
minimize the number of super$^i$-steps.

The utility of the model is demonstrated through the design of
collective communication algorithms such as gather, scatter,
reduction, prefix sums, one-to-all broadcast and all-to-all
broadcast. Two simple design principles are used, i.e. the root of
a communication operation must be a fast node and faster nodes
receive more data than the slower nodes. To validate the
predictions of the HBSP$^k$ two experiments were carried out for
both designs. It was found that not all algorithms benefits on a
heterogeneous environment. For example, broadcast (one-to-all and
all-to-all) algorithm developed using the two design principles
shows negligible benefits. The predicted and actual values for
one-to-all-broadcast communication are shown in
\reftable{one_to_all_broadcast1} and
\reftable{one_to_all_broadcast2} respectively. $p$ is the number
of processors, $T_s$ and $T_f$ denote the execution time assuming
a slow and a fast root node, respectively. $T_b$ is the runtime
for balanced workload (each node has same the amount of workload).
This is because a broadcast requires each machine to possess all
of the data elements at the end of the operation and clearly
slowest machine effects the overall performance. Thus the
conclusion driven was, any collective operation that require nodes
to possess all of the data items at the end of operations will not
be able to exploit heterogeneity.

The plus point for this model is that HBSP$^k$ gives a single
system image of a heterogeneous platform by incorporating salient
features of the underlying machines (characterized by a few
parameters). This keeps an application developer away from
non-uniformity of the underlying architecture. The model however
does not include fault tolerance issues. Some of the parameters
used are assumed to be constant, but this is not the case for
heterogeneous machines that are distributed geographically apart.
Communication between nodes depend on the network conditions,
furthermore the load of processing nodes are not constant on
Grids.

\begin{table}[htbp]
  \centering
  \caption{Table shows the predicted values for the one-to-all broadcast communication using the $HBSP^k$
  model. } 
  \label{one_to_all_broadcast1}
  \begin{tabular}{|c||cccccccccc|}
    \hline
    &\multicolumn{10}{|c|}{problem size (in KBs)} \\
    \hline
     & $100$ & $200$ & $300$ & $400$ & $500$ & $600$ & $700$ & $800$ & $900$ & $1000$  \\
    \hline
    $p=10$ &  &  &  &  &  &  &  &  &  &  \\
    $T_s$ & $0.238$ &$0.402$&$0.566$ & $0.729$ & $0.893$ & $1.057$ & $1.221$ & $1.385$ & $1.549$ & $1.712$ \\
    $T_f$ &$0.176$  &$0.278$ & $0.380$ & $0.482$ & $0.584$ & $0.686$ & $0.788$ & $0.890$ & $0.992$ & $1.094$ \\
    $T_b$ & $0.176$  & $0.278$ & $0.380$ & $0.482$ & $0.584$ & $0.686$ & $0.788$ & $0.890$ & $0.992$ & $1.094$ \\
    \hline
  \end{tabular}
  \centering
  \caption{Table shows the actual execution time for the one-to-all-broadcast communication using the $HBSP^k$
  model. } 
  \label{one_to_all_broadcast2}
  \begin{tabular}{|c||cccccccccc|}
    \hline
    &\multicolumn{10}{|c|}{problem size (in KBs)} \\
    \hline
     & $100$ & $200$ & $300$ & $400$ & $500$ & $600$ & $700$ & $800$ & $900$ & $1000$\\
    \hline
    $p=10$ &  &  &  &  &  &  &  &  &  &  \\
    $T_s$ & $1.426$ &$1.769$ & $1.452$ & $1.770$ & $2.310$ & $3.588$ & $3.332$ & $3.877$ & $4.489$ & $5.061$ \\
    $T_f$ & $0.450$ &$0.862$ & $1.266$ & $1.537$ & $2.041$ & $2.435$ & $3.152$ & $3.573$ & $4.212$ & $4.773$ \\
    $T_b$ & $0.410$ &$1.13$ & $1.134$ & $1.766$ & $1.839$ & $2.676$ & $3.269$ & $3.633$ & $4.476$ & $4.952$ \\
    \hline
  \end{tabular}
\end{table}

\subsubsection{Bulk Synchronous Parallel-GRID (BSPGRID)}
BSPGRID~\cite{Vasilev03} is a model based on BSP model for grid
based parallel algorithms. It extends the Bulk Synchronous
Parallel Random Access Machine (BSPRAM)~\cite{Tiskin98} model
which is an extension of BSP model with shared memory to reduce
the complexity involved in algorithm design and programming. A
BSPGRID is a collection of processor with limited memory units, a
shared memory with unlimited capacity, and a global
synchronization mechanism. The shared memory is likely to be a
collection of disk units in this model. At the end of each
supersteps processors are globally synchronized and the contents
of all local memories are discarded. This is in contrast with BSP
model where there is a persistency of data at processor nodes
between supersteps. The concept of virtual processors is used when
the problem size is larger than memory capacity at the processing
nodes. This implies that each physical processing units may be
required to perform work of multiple virtual processors
sequentially in a particular superstep. Processor reliability and
availability is taken into consideration by allowing the number of
physical processor to vary between supersteps. A recovery protocol
is also provided in case processors fail unexpectedly during
supersteps. An additional synchronization barrier is introduced
and the work of failed processors is rescheduled after the
barrier. It is not mentioned how the implementation of shared
memory will be done. However, a centralized shared memory
implementation would cause communication bottleneck at the master
processor, thus a likely solution is to implement virtual shared
memory distributed over the grid~\cite{Martin04}. The BSPGRID cost
model has four parameters as shown in \reftable{bspgrid}. The
model allows time and work cost to be predicted for an algorithm.
The time cost is defined as the best performance that can be
achieved if enough processors are used to solve a problem. The
work cost is defined as the processor-time product of the
algorithm.

\begin{longtable}[htbp]{p{1.0in}p{4.5in}}
\caption{ Parameters used in BSPGRID model. } 
\label{bspgrid}\\
\hline \\
Parameters & Description \\
\hline \\
$M$ & the amount of memory per processor in words. \\
$g$ & the cost of shared memory access per word. \\
$l$ & the cost of synchronization. \\
$N$ & the problem size in words. \\
\hline \\
\end{longtable}

The time cost of a superstep is defined to be:
\begin{equation}
T=w+gh+l. \nonumber
\end{equation}
where $w$=$max_iw_i$, $h$ = max $h_{i}^{in}$+ max $h_{i}^{out}$,
$w_i$, is defined as the cost of computation on processor $i$,
$h_{i}^{in}$, is the number of words read from the shared memory
to processing unit $i$, $h_{i}^{out}$, is the number of words
written to shared memory from unit $i$. The work cost of a
superstep is defined to be:
\begin{equation}
W=\upsilon T. \nonumber
\end{equation}
where, $\upsilon$, is the number of processors used during the
superstep. It is noteworthy that these costs are similar to the
PRAM model. The cost of an algorithm is taken as sum of the costs
of all of its supersteps. The unit of the cost model is taken as
the cost of a single computational operation. The value of $g$ and
$l$ are normalized to this unit. A BSPGRID computer is defined as
$BSPGRID(M,g,l)$ with fixed parameters  $M$, $g$ and $l$. The
number of processing unit is fixed and this number is derived from
the value of $M$, $N$ and the algorithm used. Execution time of an
algorithm with time cost $t$ and work cost $c$ on a $p$ processor
machine that can emulate BSPGRID machine is given by
$T(p)=(c-t)/p+t$. Computational complexity for matrix
multiplication on grid was derived using this model. This model
does not take into consideration the network and processing units
heterogeneity which is an important aspect of Grid.

\subsubsection{Dynamic BSP}

This model is a modification of BSPGRID and it addresses the
heterogeneity issues, fault tolerance and also provides the
ability to spawn additional processes within supersteps when it is
required.

\begin{figure}[htbp]\centerline{
{\includegraphics[height=3.0in,width=4.0in]{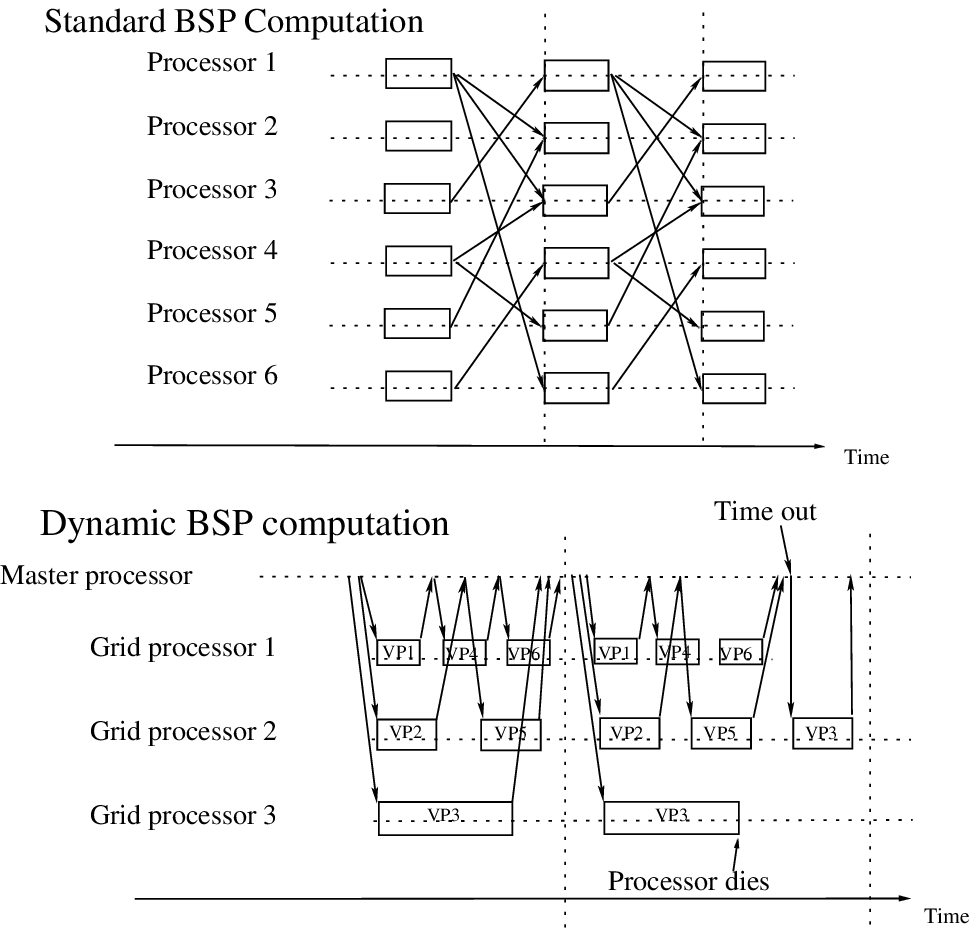}}}
\caption{The difference between standard BSP computation and
Dynamic BSP computation. } 
\label{dynamicBSP}
\end{figure}

Dynamic BSP~\cite{Martin04} uses task-farm model to implement BSP
supersteps, where individual tasks are represented as virtual
processors. The data bottleneck problem of task-farm model is
countered by using a master processor known as task server, worker
processors and a data server(implemented either as a distributed
shared memory or remote/external memory). \reffig{dynamicBSP}
shows the difference between BSP computation and the Dynamic BSP
computation. The master processor deals with task scheduling,
memory management, and resource management. At the beginning of
each superstep the master processor distributes a virtual
processor number to each physical processor.

This virtual processor in turn fetches local data from data
server, performs computations, write the output to the data server
and informs the master processor that it has finished the task.
The master processor which maintains a queue of pending virtual
processors dynamically assigns them to waiting physical
processors. When all the virtual processor have been executed in a
particular superstep, the global shared memory is restored to a
consistent state and the next superstep commences. The task farm
approach used in this model hides heterogeneity across the grid by
choosing the number of virtual processor to far exceed those of
the physical processors (this approach is known as parallel
slackness).

Fault tolerance is dealt by using timeouts, when time has exceeded
the timeout period, the physical processor is considered to have
died and the work is reallocated to another physical processor
within the same superstep as shown in \reffig{dynamicBSP}. This
model also allows the virtual processors to spawn other virtual
processors (child process). However, the child creation process
has to be registered at the master processor, where the virtual
process sends a message to master requesting it to spawn one or
more children. The standard cost model for BSP is said to be
suitable for dynamic BSP even though the value of parameters $g$
and $l$ will vary significantly between grid nodes. The author
claims that using task-farm approach together with the use of
parallel slackness would make it reasonable to utilize the
measured values for $g$ and $l$ (suitably averaged) for predicting
cost.

\subsubsection{Parameterized LogP (P-logP)} 
\begin{figure*}[htbp]
\centerline{\psfrag{g(m)}{$g(m)$}\psfrag{=s(m)}{$=s(m)$}
\psfrag{sender}{sender}\psfrag{os(m)}{$os(m)$}\psfrag{time}{time}
\psfrag{or(m)}{$or(m)$}\psfrag{time}{time}\psfrag{receiver}{receiver}
\psfrag{L}{$L$}\psfrag{+}{$+$}\psfrag{g(m)}{$g(m)$}
\psfrag{=r(m)}{$=r(m)$}
{\includegraphics[height=2in,width=3.3in]{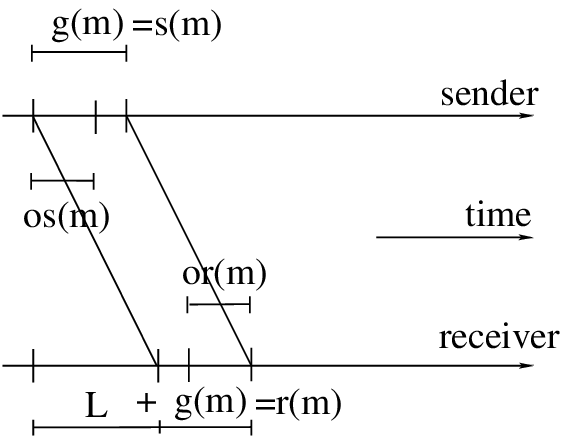}}}
\caption{\label{PLogP_fig}Message transmission in parameterized
LogP. } 
\end{figure*}

\begin{figure}[htbp]
\centerline{\psfrag{Nodes}{Nodes}\psfrag{Gateway}{Gateway}
\psfrag{LAN}{LAN}\psfrag{WAN}{WAN}
{\includegraphics[height=2in,width=3.3in]{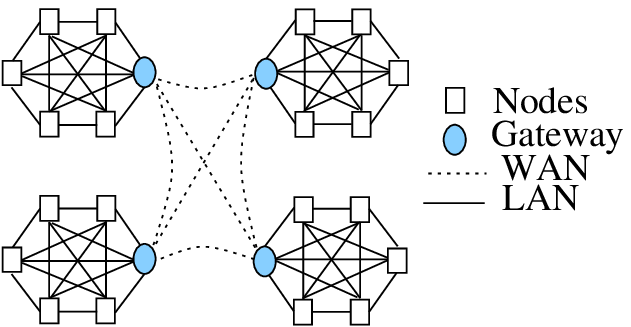}}}
\caption{\label{LAN_WAN}Clustered wide area system. } 
\end{figure}

The parameterized LogP
(P-LogP) model~\cite{Kielmann01} is an extension from
LogP~\cite{13} and LogGP~\cite{alexandrov97} model to accurately
estimate the completion time of collective communication on a wide
area systems (hierarchical systems). The existing models such as
LogP model are inaccurate for collective communication on
hierarchical systems with fast local networks and slow wide-area
networks. This is because they use constant values for overhead
and gap, also LogP is restricted to short messages while LogGP
adds the gap per byte for long messages, assuming linear behavior.
Both this models do not handle overhead for medium sized to long
messages correctly and do not model hierarchical networks. The
P-LogP model uses different sets of parameters for both networks,
and consists of five parameters as shown in \reftable{Plog_P}.
This model uses parameters as a function of message size and uses
measured values as input. A network $N$ is characterized as
$N=(L,os,or,g,P)$. The Gap parameter, $g(m)$ is also known as the
reciprocal value of the end-to-end bandwidth from process to
process for messages of size $m$. This parameter models the time a
message ``occupies" the network, as such the next message cannot
be sent before $g(m)$ time. Hence, $r(m)=L+g(m)$ is the time the
receiver has received the message. The latency $L$ on the other
hand can be viewed as time taken for the first bit of message to
travel from sender to receiver. This model is depicted in
\reffig{PLogP_fig}, values of these parameters are obtained from
empirical studies.
\begin{center}
\begin{longtable}[htbp]{p{0.6in}p{4.5in}}
\caption{\label{Plog_P} Parameters used in P-LogP model. } 
 \\
\hline \\
Parameters & Description \\
\hline \\
$P$ & Number of processors. \\
$L$ & End-to-end latency from process to process (it combines all
contributing factors such as copying data to
and from network interfaces and transfer over the physical network). \\
$os(m)$ & Send overhead (time the CPUs are busy sending messages as a function of message size). \\
$or(m)$ & Receive overhead (time the CPUs are busy receiving messages as a function of message size). \\
$g(m)$ & Gap (minimum time interval between consecutive message
transmissions or receptions along the same link or connection as a function of message size). \\
\hline \\
\end{longtable}
\end{center}
When a sender sends multiple messages in a row, the latency cost
contributes only once to the completion time but the gap values of
all messages sum up as,
$r(m_1,\ldots,m_n)=L+g(m_1)+\ldots+g(m_n)$.
For clustered wide area systems, two parameter sets are used, i.e
for LAN and WAN with subscript $l$ and $w$ respectively. For a
local area network, the time taken for the receiver to receive the
message is given by: $r_l(m)=L_l+g_l(m)$ and the time taken for
sending a message of size $m$ is given by: $s_l(m)=g_l(m)$. For
wide area transmission, there are three steps: the sender sends
message to its gateway, this gateway sends the message to the
receiver's gateway and finally the receiver's gateway sends the
message to the receiving node, refer \reffig{LAN_WAN}. The value
of $r_w$ depends on wide area bandwidth and is expressed as an
analogy to $r_l$. Value of $s_w$ is determined by wide-area
overhead $os_w(m)$ or local-area gap $g_l(m)$, whichever is
higher. Thus the equations for wide-area case is:
$s_w(m)=max(g_l(m),os_w(m))$ and $r_w(m)=L_w+g_w(m)$.

Performance model for single layer broadcast algorithm is given as
$T=(k-1)\cdot\gamma(m)+\lambda(m)$ for $k$ message segment of size
$m$. Here, latency $\lambda(m)$ and gap $\gamma(m)$ is of a
broadcast tree analogous to $L$ and $g(m)$ for a single message
send. $\lambda(m)$ denotes time taken for message to be received
by all nodes, after root process starts sending it. $\gamma(m)$ is
the time interval between the sending of two consecutive segments
(indicates the throughput of a broadcast tree). For example values
of $\gamma(m)$ and $\lambda(m)$ for flat WAN tree used in
MagPIe~\cite{Kielmann99} is:
\begin{equation}
\gamma(m)=\text{max}(g(m),(P-1)\cdot s(m)), \nonumber
\end{equation}
\begin{equation}
\lambda(m)=(P-2)\cdot s(m)+r(m). \nonumber
\end{equation}

Here, $\lambda(m)$ is the maximum of the gap between two segments
of size $m$ sent on the same link and the time the root needs for
sending $(P-1)$ times the same segment on disjoint links. The
corresponding value for $\lambda(m)$ is the time at which a
message segment is sent to the last node, plus the time it is
received.

For general tree shape, upper bounds for both parameters can be
expressed depending on the degree $d$ and height $h$ of a
broadcast tree:
\begin{equation}
\gamma(m)\leq \text{max}(g(m),or(m)+d\cdot s(m)), \nonumber
\end{equation}
\begin{equation}
\lambda(m) \leq ((d-1)\cdot s(m)+r(m)). \nonumber
\end{equation}
Here, $\lambda(m)$ is the maximum of the gap caused by the
network, and the time a node needs to process the message. For
intermediate nodes, this is the time to receive the message plus
the time to forward it to $d$ successor nodes (for the root and
for the leaf nodes, it is either one of both).The exact value of
$\lambda(m)$ depends on the order in which the root process and
all intermediate nodes send to their successor nodes and which
path leads to the node that receives the message last.

P-LogP model is used to optimize four type of collective
communications, namely broadcast, scatter, gather and all-gather
in the MagPIe~\cite{Kielmann99} message passing library.

\subsection{Summary}
In general, it is clear that all the computational models are
trying to incorporate factors that effect data movements to
accurately predict performance of parallel algorithms. A pattern
that we observe in the traditional models is that they tend to focus
on architectural parameters only rather than on both the
algorithmic and architectural parameters. On WAN, factors that
contribute to performance of inter-processor communication change
very rapidly due to shared network and shared computing resources.
As a result, it is impossible to predict performance of parallel
applications accurately. However, it is very important to have
some idea of the behavior of the WAN before a parallel application
is deployed on it. We also see that the trend in computational
models for WAN are to emphasize more towards tuning different
types of communication that is frequently used in parallel
algorithms by using empirically gathered information. This makes
sense because the main bottleneck in parallel computing over WAN
is the communication phase, assuming computational resources are
reserved (available unconditionally without any failure) in
advance for usage. There are many other factors that contribute to
the performance of parallel programs on the WAN, and it is
impossible at least at the moment to include all the factors and
find an optimal solution in real time to obtain good speedup for
parallel applications. It is also worth noting that the use of
stochastic approach for computational models may be inevitable
because of the unpredictable nature of the computational resources
and the WAN.

\section{Programming Libraries}\label{sec6}
Programming libraries play a very significant role in simplifying
complexity involved in writing parallel programs. These libraries
provide frequently used commands for developing parallel
applications on HPC architectures. Historically, the main focus of
programming language development has been on expressibility, and
providing constructs which translate and preserve algorithmic
intentions. However, lately the focus of language development has
begun to include performance issue in addition to
expressibility~\cite{6}. Performance issues are usually related to
efficiently moving data. The cost of moving data between memory or
storage to processing units and between processing units usually
contributes considerably to the total computation time. In order
to reduce this cost, many new algorithms (e.g. for collective
communication) uses performance model to assist in tuning the
parameters used for the communication~\cite{Kielmann99}.

In this section, we study some parallel programming libraries
commonly used for parallel computing in System Area Network (SAN),
Local Area Networks (LAN) and Wide Area Networks (WAN).

\subsection{Parallel Virtual Machine (PVM)}
PVM is a set of software tools and libraries that emulates  a
general-purpose, flexible, heterogeneous computing framework on an
interconnected computers of varied
architecture~\footnote{http://www.netlib.org/pvm3/book/node17.html}.
The system is composed of two parts: 1) A daemon, called pvmd3
that resides on all computing nodes which makes up the virtual
machine. Daemon can run on heterogeneous distributed computing
nodes connected by different type of network topology. 2) An API
that contains a library of PVM interface routines required to
communicate between processes in an application. Processes can
interact between each other via message passing where messages are
send to and received using unique ``task identifiers" (TIDs) which
are the identifier for all PVM tasks in a parallel application.
PVM supports C, C++ and Fortran languages.~\cite{Geist94}

\subsection{Message Passing Interface (MPI)}
Message Passing Interface
(MPI)~\footnote{http://www.cs.usfca.edu/mpi/} is a successful
community standard for the extended portable message passing model
of parallel communications. MPI is a specification and not a
particular implementation. There are many implementation of MPI
such as MPICH, LAM/MPI (runs on networks of Unix/Posix
workstations), MP-MPICH (runs on Unix systems, Windows NT and
Windows 2000/XP Professional), WMPI runs on Windows platform and
MacMPI (MPI implementation for Macintosh computers). A more
complete list of MPI implementation is available at LAM
website~\footnote{http://www.lam-mpi.org/mpi/implementations/fulllist.php}.
The most popular parallel implementation of these is the MPICH
from Argonne National Laboratory. A correct MPI program should be
able to run on all MPI implementation without change. The standard
includes point-to-point communication, collective communication,
process groups, communication contexts, process topologies,
environmental management and inquiry, bindings for Fortran77 and C
and also profiling interface. In message passing model each
process executing in parallel have separate address spaces. It
however does not include explicit shared-memory operations;
operations that require more operating system support than is
currently standard: e.g. interrupt-driven receives, remote
execution, or active messages;program construction tools;
debugging facilities; explicit support for threads; support for
task management; and I/O functions~\cite{Gropp99}.

\subsection{Paderborn University BSP (PUB)}
The Paderborn University BSP library is a C communication library
based on BSP model. This implementation supports buffered as well
as unbuffered non-blocking communication between any pair of
processors. It also provides nonblocking collective communication
operation such as broadcast, reduce and scan on any arbitrary
subsets of processors. These primitives are however not available
on Oxford BSP toolset or Green BSP library. Another different
aspect of PUB is the possibility to dynamically partition the
processors into independent subsets. As such PUB allows support
for nested parallelism and subset synchronization. PUB also
supports a zero-cost synchronization mechanism known as oblivious
synchronization. The concept of BSP objects is introduced in PUB
which serve three purposes. They are used to distinguish the
different processor groups that exist after a partition operation,
for modularity and safety purposes and can be used to ensure that
messages sent in different threads do not interfere with each
other and that a barrier synchronization executed in one thread
does not suspend the other threads running on the same
processors~\cite{56}. The most useful feature of BSP library
variants compared to other model is the ability to construct a
cost function using BSP parameters (p,r,g,l) which represents
number of processors, computing rate , communication cost per data
word and global synchronization cost respectively to predict
performance and scalability of parallel programs. Other
programming libraries that are conceptually based on BSP model
include BSPlib~\cite{hill98}, Green BSP~\cite{Goudreau96},
xBSP~\cite{Kee01}, and BSPedupack~\cite{Bisseling04}.

\subsection{MPICH-G2}
MPICH-G2~\cite{Karonis03,Foster98} is a grid enabled
implementation of the Message Passing Interface (MPI) that allows
a user to run MPI programs across multiple computers at different
sites using the same commands that would be used on a parallel
computer. This library extends the Argonne MPICH implementation of
MPI to use services provided by the Globus grid toolkit for
authentication, authorization, resource allocation, executable
staging, and I/O as well as process creation, monitoring, and
control. Various performance critical operations, including
startup and collective communication, are configured to exploit
network topology information. The library also exploits MPI
constructs for performance management, e.g., the MPI communicator
construct is used for application-level discovery of both network
topology and network quality-of-service. Adaptation is then
performed for both the information. The major difference between
MPICH-G2 and its predecessor MPICH-G is that the Nexus component
which provided the communication infrastructure has been removed.
The MPICH-G2 now handles communication directly by re-implementing
Nexus with other improvements. This improvements include increased
bandwidth, reduced latency for intra-machine, more efficient use of
sockets, support for MPI\_LONG\_LONG and MPI-2 file operations and
added C++ support.

\subsection{PArallel Computer eXtension (PACX MPI)}
The PACX-MPI~\cite{Beisel97,Gabriel98} library enables parallel
applications to seamlessly run on a computational grid such as
cluster of MPPs connected through high speed high-speed networks
or even the Internet. Among the goal of this programming library
is to provide users with a single virtual machine, run MPI
programs without any modification on computational grid, use
highly tuned MPI for internal communication on each participating
MPP, and use fast communication for external
communication.~\cite{Beisel97}

\subsection{Seamless thinking aid MPI (StaMPI)}
StaMPI~\cite{Tsujita02} is the application-layer communication
interface for the Seamless Thinking Aid from JAERI (Japan Atomic
Energy Research Institute). It is a meta-scheduling method which
includes MPI-2 features to dynamically assign macro-tasks to
heterogeneous computers using dynamic resource information and
static compile time information. StaMPI automatically chooses
vendor specific communication library for internal communication
between processors and Internet Protocol (IP) for external
communication between processor on different parallel computers.
It also facilitates automatic message routing process to enable
indirect communication between processes on different parallel
computers if these processes cannot communicate directly through
IP.

\subsection{MagPIe}
MagPIe~\footnote{http://www.cs.vu.nl/albatross/} is an optimized
collective communication library for wide area systems based on
the widely use MPI implementation, MPICH. It is available as a
plug-in to MPICH. The new collective communication algorithms used
in this library sends minimal amount of data over the slow wide
area links, and only incur a single wide area latency and it also
takes into consideration the hierarchical structure of the network
topology into account. In addition to basic send and receive there
are fourteen different collective communication operation defined.
Programmers are free to use any programming model and the details
of wide area system are hidden completely to reduce parallel
programming complexity. The wide area algorithms design were based
extensively on two conditions: 1) Every sender-receiver path used
by an algorithm contains at most one wide area link. 2) No data
items travels multiple times to same cluster. Condition (1)
ensures wide area latency contributes at most once to an
operation's completion time and condition (2) prevents wastage of
precious wide area bandwidth. Results from ~\cite{Bernaschi98},
suggests that different performance characteristics of local area
and wide area links dictate different communication graphs for
local area and wide area traffic. This has lead to two different
types of graphs being introduced: an intra cluster graph that
connects all processors within a single cluster and an inter
cluster graph that connects the different clusters. A coordinator
node is designated within each cluster to interface both the
graphs~\cite{Kielmann99}.

\subsection{Summary}
Parallel programming libraries provide many functions that are
frequently used to develop parallel applications. Functions such
as initiating socket connections, opening ports for communication,
providing secure communication between nodes, performing
collective communications using a suitable algorithm depending on
message sizes can all be performed seamlessly by using these 
libraries. More recent versions of parallel programming 
libraries which are usually an extension of existing
programming libraries tend to include information about network
condition, providing fault tolerance, adding checkpointing and
migration to better accommodate the dynamics and unreliability of
computational resources distributed geographically
apart~\cite{dikken94,casas95,Geist97,Bouteiller03}.

\section{Conclusions}\label{sec7}

The role of a parallel processing model is to show the complexity of a parallel
algorithm on a given architecture so that application developers can gauge the
performance of their application as they scale it up in size and also make
decisions concerning which resources to improve in order to increase
performance further. In this survey we have covered the problems, architectures
and models that are available for this purpose. We also covered the supporting
programming libraries, tools and utilities. It is clear that architectures are
tending towards use of commodity resources and that computational models that
describe these architectures have not become advanced enough to allow general
parallel computing in these new architectures. Hence we see embarassingly
parallel, data parallel and parametric algorithms as predominant examples of
successful deployments and utilities such as MPICH-G being used only when 
message passing is required over a wide area.

HPC architecture components such as processor speed, memory, storage,
memory-processor bandwidth, interprocessor communication bandwidth, and number
of processors used have all improved significantly over the years. However,
developing efficient parallel applications on these significantly more powerful
architectures has also increasingly become more difficult due to both the
application's and the architecture's complexity. Computational models were
developed for traditional and conventional architectures and some are becoming
available for contemporary architectures but none appear to have become widely
acceptable.

Computational models play an important role in producing efficient parallel
applications. A good model should: 1) consider characteristics of the problem;
2) consider properties of the architecture; and 3) provide important
information for programmers to translate the problem into an efficient parallel
program. Many models have been developed for traditional parallel
architectures, however it can be concluded that, it may not be possible to use
a single model to represent all the architectures because of the diversity in
application requirements and architecture heterogeneity. The other constraint
in developing good computational models is to accurately reflect data movement
between different levels of memory, storage and processors. The bandwidth
capacity, latency and communication patterns for distributing data from one
location to another have significant impact on performance and efficiency of a
parallel program.

On dedicated HPC architectures, architectural parameters that contribute to
performance of moving data such as bandwidth and latency, are usually
predictable accurately. However, on a shared environment such as a grid these
parameters are always dynamic hence contributing to inaccuracy in performance
prediction. In the past this has been attributed to: 1) Fast pace of
architectural development; 2) empirical data is often required and is too
specific to the computing environment; 3) change in resource availability for
computation due to many different processes running concurrently; 4)
uncertainty in the communication performance because of unpredictable internet
behavior.  \reftable{param_for_grid} lists the computational and communication
parameters that can effect performance of parallel algorithms on grids.

\begin{center}
\footnotesize{
\begin{longtable}{p{2.5in} p{3.5in}}
\caption{\label{param_for_grid}Computational and communication
characteristics that should be considered for the Grid
environment.}\\
 \hline \\
\textbf{Computation.} & \textbf{Communication.} \\
\hline \\
\textbf{Processor.} & \textbf{Type of interconnect.} \\
\checkmark \emph{Clock speed,} & \checkmark \emph{Network interface,} \\
\checkmark \emph{Architecture type $32$/$64$ bit,} & \checkmark \emph{LAN interconnect.} \\
    \checkmark \emph{Single or multi-core chip,} & \textbf{Communication protocol.}\\
    \checkmark \emph{CPU utilization,} &  \checkmark \emph{UDP/TCP}\\
    \checkmark \emph{No. of processors.} & \textbf{Application communication patterns/characteristics.}\\
\textbf{Memory hierarchy (L1, L2 \& L3).} & \checkmark \emph{All-to-all, gather, scatter, all-gather, broadcast etc.}\\
    \checkmark \emph{size of cache per chip,} &  \textbf{Network tuning parameter.}\\
    \checkmark \emph{size of byte line,} & \checkmark \emph{packet size,round trip time, hops, bandwidth and latency.}\\
    \checkmark \emph{size of associative way,} & \textbf{Competing network traffic.}\\
    \checkmark \emph{bandwidth between cache level.} & \textbf{Interprocessor communication
    bandwidth.}\\
    \checkmark Main memory. & \textbf{Synchronization.}\\
    $\star$ \emph{size,} & \textbf{Storage.}\\
    $\star$ \emph{utilization,} & \checkmark \emph{connectivity of disk to node (consists of many cpus)}\\
    $\star$ \emph{cpu-memory bandwidth,} & \checkmark \emph{filesystem bandwidth}\\
    $\star$ \emph{block memory transfer,} & \checkmark \emph{disk speed}\\
    & \checkmark \emph{size of storage,}\\
    & \checkmark \emph{type of filesystem,}\\
    & \checkmark \emph{storage-memory bandwidth.}\\
    \hline \\
\end{longtable} }
\end{center}

Other issues that are outside the scope of this paper but that can be
considered include fault tolerance, adaptability/autonomity, work flow and
other HPC research such as scheduling, and super-scheduling.

\bibliography{MPCG}
\bibliographystyle{plain}
\end{document}